\PassOptionsToPackage{sort&compress}{natbib}
\documentclass[5p,times]{elsarticle}

\usepackage{float} 
\usepackage{placeins}

\usepackage{enumitem}
\usepackage{kantlipsum}
\usepackage{stfloats}
\usepackage{xurl}
\usepackage{caption}
\usepackage{graphics}
\usepackage{verbatim}
\usepackage{pifont}

\usepackage{amsmath}
\usepackage{cancel}
\usepackage{graphicx}
\usepackage{wrapfig}

\usepackage{color} 
\usepackage[export]{adjustbox}
\usepackage{colortbl}
\usepackage{algpseudocode}
\usepackage{verbatim}
\usepackage{lipsum}
\usepackage{tikz}
\usepackage{enumitem,kantlipsum}
\usepackage{textcomp}
\usepackage{stfloats}
\usepackage{xurl}
\usepackage{caption}
\usepackage{graphics}
\usepackage{tabularx}
\usepackage{xcolor}
\usepackage{forest}
\usepackage{microtype}
\usepackage{graphicx}
\usepackage{longtable}
 
\usepackage{bbding}
\usepackage{afterpage}
\usepackage{etoolbox}
\usepackage{graphicx}

\usepackage{lscape}
\usepackage{ltxtable}
\usepackage{hhline}
\usepackage{caption}
\usepackage{subcaption}
\usepackage{algorithm} 
\usepackage{color} 
\usepackage{colortbl}
\usepackage{algpseudocode}
\usepackage{verbatim}
\usepackage{lipsum}
\usetikzlibrary{automata, arrows.meta, positioning}
\usetikzlibrary{decorations.pathreplacing}
\usepackage{amsthm}
\usepackage{booktabs}
\usepackage[flushleft]{threeparttable}
\usepackage{algorithm}
\usepackage{inputenc}
\usepackage{adjustbox}
\usepackage{url} 
\usepackage{multirow}
\usepackage{color,soul}
\usepackage{lipsum}
\usepackage{mathtools}
\usepackage{cuted}
\usepackage{xcolor}
\usepackage{microtype}
\usepackage{mdframed}
\usepackage{cancel}
\usepackage{wasysym}
\usepackage{pgfplotstable}
\usetikzlibrary{arrows.meta}
\usepackage{smartdiagram}
\usetikzlibrary{positioning, fit, calc, shapes, arrows,trees}
\usepackage[normalem]{ulem}
\definecolor{lemon}{rgb}{1.0, 1.0, 0.13}

\definecolor{columbiablue}{rgb}{0.61, 0.87, 1.0}

\tikzset{%
thick arrow/.style={
-{Triangle[angle=120:1pt 1]},
line width=0.6cm, 
draw=blue!20
},
arrow label/.style={
text=black,
align=center
},
set mark/.style={
insert path={
node [midway, arrow label, node contents=#1]
}
}
}
\newcommand\deleted{\bgroup\markoverwith{\textcolor{red}{\rule[0.5ex]{2pt}{0.4pt}}}\ULon}
 
\newcommand\doublecheck{\textcolor{black}{\checkmark\kern-0em\checkmark}}
\newcommand\semidoublecheck{\textcolor{black}{\checkmark\kern-0em\bcancel{\checkmark}}}

\usetikzlibrary{pgfplots.groupplots}

\definecolor{BulletsColor}{rgb}{0, 0, 0.9}
\newlist{myBullets}{itemize}{1}
 
\setlist[myBullets]{
label={\textbullet},
leftmargin=*,
topsep=0ex,
partopsep=0ex,
parsep=0ex,
itemsep=0ex,
}
 
\usepackage{cuted}
\usetikzlibrary{arrows.meta}
\usepackage{smartdiagram}
\usetikzlibrary{positioning, fit, calc, shapes, arrows,trees}
 
\definecolor{columbiablue}{rgb}{0.61, 0.87, 1.0}
 
\tikzset{%
thick arrow/.style={
-{Triangle[angle=120:1pt 1]},
line width=0.8cm, 
draw=blue!20
},
arrow label/.style={
text=black,
align=center
},
set mark/.style={
insert path={
node [midway, arrow label, node contents=#1]
}
}
}

\begin{document}
 
\begin{frontmatter}
\title{Security Threat Modeling for Emerging AI-Agent Protocols: A Comparative Analysis of MCP, A2A, Agora, and ANP
}
 
\author[First]{Zeynab Anbiaee\corref{cor}}
\ead{saba.anbiaee@unb.ca}
\author[First]{Mahdi Rabbani}
\ead{m.rabbani@unb.ca}
\author[Second]{Mansur Mirani}
\ead{mansur.mirani@mastercard.com}
\author[Second]{Gunjan Piya}
\ead{sunny.piya@mastercard.com}
\author[Second]{Igor Opushnyev}
\ead{igor.opushnyev@mastercard.com}
\author[First]{Ali Ghorbani}
\ead{ghorbani@unb.ca}
\author[First]{Sajjad Dadkhah}
\ead{sdadkhah@unb.ca}

\address[First]{Canadian Institute for Cybersecurity (CIC), University of New Brunswick, New Brunswick, Canada.}
\address[Second]{Mastercard Vancouver Tech Hub, Vancouver, British Columbia, Canada.}

\cortext[cor]{Corresponding author}

\begin{abstract}
The rapid development of the AI agent communication protocols, including the Model Context Protocol (MCP), Agent2Agent (A2A), Agora, and Agent Network Protocol (ANP), is reshaping how AI agents communicate with tools, services, and each other. While these protocols support scalable multi-agent interaction and cross-organizational interoperability, their security principles remain understudied, and standardized threat modeling is limited; no protocol-centric risk assessment framework has been established yet. This paper presents a systematic security analysis of four emerging AI agent communication protocols.
First, we develop a structured threat modeling analysis that examines protocol architectures, trust assumptions, interaction patterns, and lifecycle behaviors to identify protocol-specific and cross-protocol risk surfaces. 
Second, we introduce a qualitative risk assessment framework that identifies twelve protocol-level risks and evaluates security posture across the creation, operation, and update phases through systematic assessment of likelihood, impact, and overall protocol risk, with implications for secure deployment and future standardization.
Third, we provide a measurement-driven case study on MCP that formalizes the risk of missing mandatory identity binding validation for executable components as a falsifiable security claim by quantifying wrong-provider tool execution under multi-server composition across representative resolver policies. Collectively, our results highlight key design-induced risk surfaces and provide actionable guidance for secure deployment and future standardization of agent communication ecosystems.

\end{abstract}

 
\end{frontmatter}

%%%%%%%%%%%%%%%%%%%%%%%%%%%%%%%%%%%%%%%%%%%%%%%%%%%%%%%%

\section{Introduction}
For decades, scientists tried to make the systems intelligent; the real journey of AI started with symbolic AI and expert systems, which were rule-based and rigid \cite{xiong2024converging}. Then, Machine Learning (ML) came for enabling systems to learn patterns from data rather than relying on hard-coded rules \cite{de2025comparative}. As data and computational power grew, Deep Learning (DL) emerged, giving rise to powerful models for vision, speech, and more. That evolution led to large language models (LLMs), which understand and generate human language at scale \cite{naveed2025comprehensive, haenlein2019brief}. But now, it is time for a new phase: the age of AI agents. These are not just passive models waiting for user prompts; they are proactive and autonomous entities capable of interacting with tools, environments, and other AI agents \cite{luo2025large, park2023generative}. Providing secure and structured communication between AI agents is a foundation for what’s next: Artificial General Intelligence (AGI) and even Artificial Superintelligence (ASI) \cite{kim2024road}, where intelligent agents collaborate seamlessly in real-time environments \cite{ehtesham2025survey}.

%%%%%%%%%%%%%%%%%%%%%%%%%%%%%%%NEW%%%%%%%%%%%%%%%%%%%%%%%%%%%%%%%%%%%%%%%%

\begin{figure*}[!t]
    \centering
    \includegraphics[width=\linewidth]{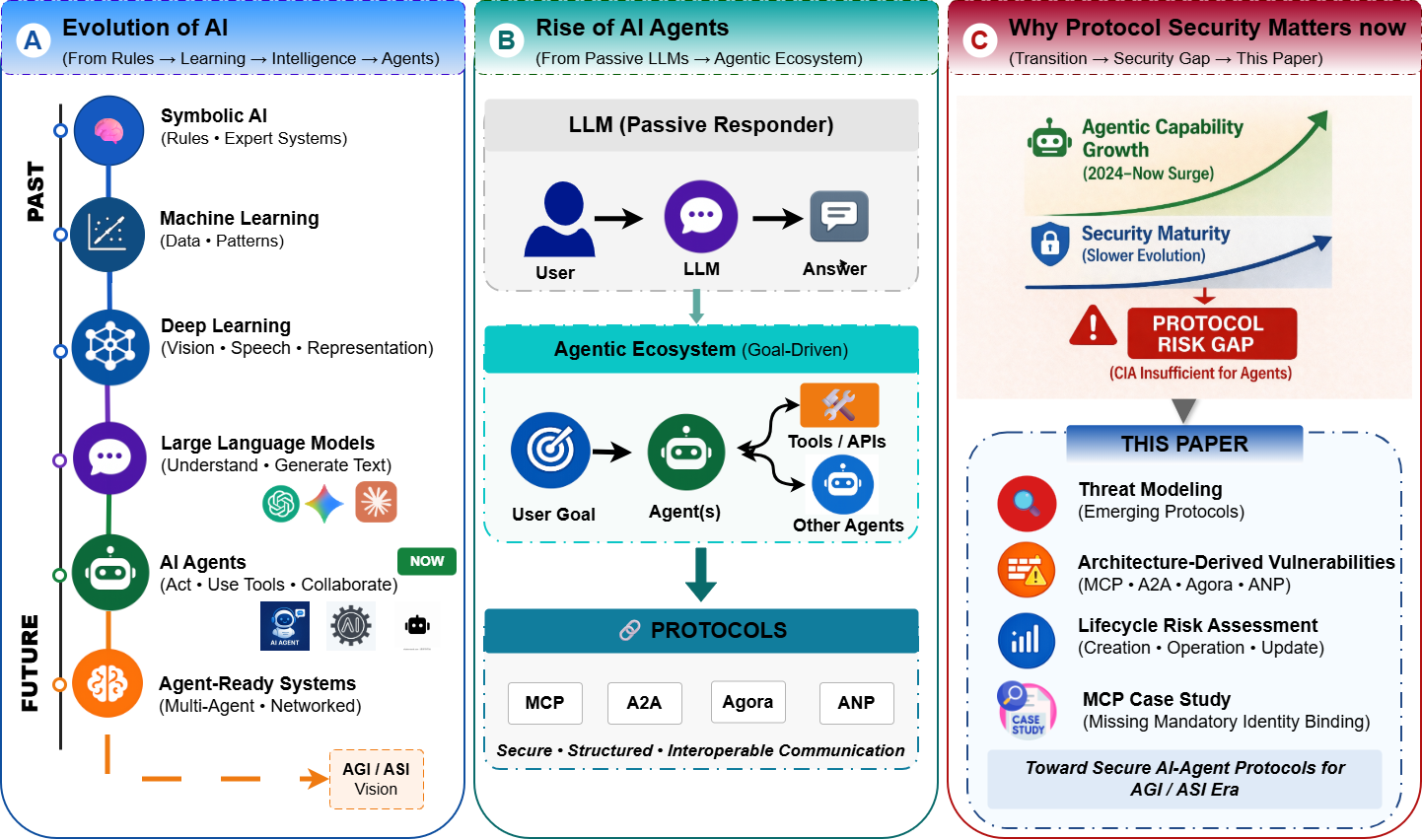}
    \caption{Evolution of AI toward agentic systems, the shift from passive to proactive interaction, and the resulting protocol security gap motivating this work.}
    \label{fig:ai_timeline}
\end{figure*} 

%%%%%%%%%%%%%%%%%%%%%%%%%%%%%%%%%%%%%%%%%%%%%%%%%%%%%%%%%%%%%%%%%%%%%%%
The rapid advancement of agentic AI in 2025 further accelerated this transition. The integration of autonomous agents with LLMs enhanced system capabilities in complex reasoning and inter-agent communication \cite{luo2025large, belcak2025small}. These agents serve as the foundation of intelligent systems to fulfill the intent of a user's prompt \cite{park2023generative}. This architecture transforms LLMs from passive responders into dynamic, goal-driven entities operating within multi-agent ecosystems \cite{luo2025large}. A broader systemic transition is expected in 2026, as more companies and organizations begin reshaping their systems to become \textit{agent-ready}. This means going beyond traditional software design to ensure that backend services can be accessed by autonomous AI agents.

From a security and privacy perspective, however, this evolution poses significant challenges, particularly given that cybersecurity mechanisms are not evolving at the same pace as agentic AI systems. Security teams are still grappling with new threats introduced by LLMs between 2023 and 2025 \cite{das2025security, li2025security}. Traditional data security frameworks centered on confidentiality, integrity, and availability (CIA) are no longer sufficient in AI agent environments. \cite{louck2025security}. Securing these complex communication processes requires a systematic investigation of vulnerabilities in the communication protocols used by agentic AI systems. Limited awareness of protocol-level interactions and dependencies directly affects exposure surfaces, trust boundaries, and failure propagation across multi-agent workflows. However, existing studies remain fragmented and largely focus on isolated protocol risks or specific implementations, without offering a unified view of vulnerability classes \cite{ehtesham2025survey, louck2025security}.

To address this gap, this paper advances the emerging area of AI-agent protocol security through a systematic and forward-looking analysis of communication risks in agentic ecosystems. Although each emerging protocol in the literature follows a different architecture and introduces its own set of vulnerabilities that require independent analysis, examining isolated weaknesses alone is not sufficient to capture system-level risks. Instead, this paper provides a protocol-centric perspective that integrates threat modeling, architectural analysis, and lifecycle-aware risk assessment across multiple emerging protocols and offers a structured basis for comparative evaluation and proactive mitigation. The selection of the four communication protocols is primarily based on two criteria: popularity and maturity \cite{yang2025survey}. Among many emerging standardization efforts, these protocols have progressed beyond conceptual proposals and have begun to see practical integration, making them suitable for systematic security analysis. The contributions of this paper are as follows:

%%%%%%%%%%%%%%%%%%%%%%%%%%%%%%%%%%%%%%%%%%%%%
% \begin{figure}[!t]
%     \centering
%     \includegraphics[width=3.2 in]{Figures/organization_survey2.pdf.pdf}
%     \caption{The organization of this paper}
%     \label{organization_paper}
% \end{figure}

%%%%%%%%%%%%%%%%%%%%%%%%%%%%%%%%%%%%%%%

\begin{itemize}
    \item We present a structured, comparative security threat modeling analysis of MCP, A2A, ANP, and Agora, consolidating scattered early findings and protocol documentation into a coherent taxonomy that highlights risk surfaces under realistic deployment assumptions.
    \item We perform an architecture-based analysis and derive a catalog of design-induced threat hypotheses for MCP, A2A, ANP, and Agora, grounded in trust boundaries, identity/authorization binding assumptions, and cross-protocol composition risks.
     \item \textcolor{black}{We present a systematic qualitative risk assessment of major AI agent communication protocols, identifying twelve protocol-level risks and proposing a lifecycle-aware framework that evaluates security posture across the creation, operation, and update phases, with implications for secure deployment and future standardization.}
     \item We contribute a measurement-driven case study that formalizes the lack of mandatory identity binding validation for executable components in MCP as a falsifiable security claim by quantifying wrong-provider tool execution under multi-server composition across multiple realistic resolver policies.  
\end{itemize}

Figure \ref{fig:ai_timeline} illustrates the evolution of AI, the shift from passive to proactive agentic interaction, and the resulting protocol security gap addressed in this work.

The rest of this paper is organized as follows. Section \ref{Section_II} reviews related work on agent communication and protocol security. Section \ref{Section_III} introduces the foundations of agent communication protocols and representative frameworks. Section \ref{Section_IV} presents the threat model and key security dimensions. %Section \ref{Section_V} analyzes potential threats arising from protocol design and system architecture. 
Section \ref{Section_VI} describes the evaluation methodology and assessment framework. Section~\ref{case_study} provides an experimental case study to demonstrate the applicability of the proposed analysis framework on real-world agent communication protocols. Section \ref{Conclusion} concludes the paper and summarizes the main findings. Finally, Section \ref{Future_work} discusses future research directions.

%%%%%%%%%%%%%%%%%%%%%%%%%%%%%%%%%%%%%%%%%%%%%%%%%%%%%%%%

\section{Related Work}
\label{Section_II}
Research on the security and privacy of AI systems has expanded rapidly with the fast deployment of LLMs \cite{hou2025unveiling}. Although concerns related to LLMs are not new, many studies in the literature have already examined the security issues associated with standalone LLM models \cite{yao2024survey}. More recently, as LLMs have been integrated with external resources through retrieval-augmented generation (RAG), these security concerns have become even more complex \cite{zeng2024good}. The focus of \cite{hughes2025ai} is on a mature form of agentic AI in which all major components are present, including the LLM, orchestration mechanisms, and multiple agents operating under coordinated control. \cite{tran2025multi} is among the first to take a broad view of ecosystems to highlight existing vulnerabilities.

In early 2025, several communication protocols were proposed to support coordination among agents \cite{Bizety2025AIProtocols}. Most of these protocols remain at a conceptual stage and have not yet been fully implemented or evaluated in real-world environments \cite{yang2025survey}. Therefore, this survey focuses on four relatively mature and most popular protocols, MCP, A2A, Agora, and ANP, for detailed analysis \cite{radosevich2025mcp}. To the best of current knowledge, this study represents one of the first efforts to conduct risk assessment and security threat modeling across agentic AI communication protocols. For comparison with related works in the literature, studies with high relevance as well as those with moderate relevance to security and privacy in AI agent communication are considered. Table \ref{survey_comparison} provides a concise comparison of existing related works and highlights how the identified gaps are addressed in this paper through risk assessment and threat prediction across the four main AI agent protocols.

%%%%%%%%%%%%%%%%%%%%%%%%%%%%%%%%%%%%%%%%%%%%%%%%%%%%%%%%%%%%%%%%%%%%%%%%%%%%

\subsection{Security Studies on MCP Communication Protocols}

In the last two years, researchers have approached security for the MCP from several angles, though none have yet delivered a fully formal threat model. Hou and colleagues \cite{hou2025model} offer the first deep dive into MCP’s architecture and operation and walk through the kinds of problems that can arise at life cycle stages.
Narajala et al. \cite{narajala2025enterprise} constructed a defense-in-depth security framework under a zero-trust assumption. Ehtesham et al. \cite{ehtesham2025survey} present a comparative analysis and roadmap for AI agent communication protocols rather than proposing new security mechanisms or formally assessing attack surfaces. Yang et al. \cite{yang2025iot} integrate MCP servers in IoT environments and evaluate protocol performance, but provide little discussion of security. Conversely, few design-oriented studies focus on protecting integrity to detect tampering with tools and servers \cite{yang2025iot}, \cite{li2025secure}. Taken together, the field still lacks formal adversary models, empirical attack-based evaluations, and cross-protocol analysis, leaving substantial scope for more rigorous and systematic security research on MCP.

%%%%%%%%%%%%%%%%%%%%%%%%%%%%%%%%%%%%%%%%%%%%%%%%%
\subsection{Security Studies on A2A Communication Protocols}

A case study by Duan and Lu \cite{duan2025agent} assesses the A2A protocol for edge-computing environments, noting that edge-based multi-agent systems face heterogeneity, scalability, dynamicity, and resource constraints. Habler et al. \cite{habler2025building} analyze the security of A2A using the MAESTRO threat-modeling framework, with emphasis on vulnerabilities in Agent Card management, task-execution integrity, and authentication. Louck et al. \cite{louck2025proposal} provide recommendations to enhance the A2A protocol for handling sensitive data in multi-agent workflows. The authors propose protocol-level refinements; however, these remain largely conceptual and example-driven.

A broad multi-agent security study by He et al. \cite{he2025comprehensive} shows that vulnerabilities in individual components can cascade through inter-agent communications. A2A has also been extended to specific domains: Duan et al. \cite{duan2025ai} propose an AI‑Agent Communication Network (ACN) for 6G environments where security is treated as an inherent property of the underlying network rather than being addressed at the protocol level. In summary, these studies underscore that while A2A provides a foundational communication framework, robust security requires formal threat models, cryptographic enhancements, and empirical validation.

%%%%%%%%%%%%%%%%%%%%%%%%%%%%%%%%%%%%%%%%%%%%%%%%%%%%%%%%%%%%%%%%%%%%%%%%%%%%%%%%%%%%new table for related works%%%%%%%
\newcommand{\cmark}{\textcolor{green!60!black}{\ding{51}}}
\newcommand{\xmark}{\textcolor{red}{\ding{55}}}

\begin{table*}[t]
\centering
\scriptsize
\setlength{\tabcolsep}{3pt}
\caption{\textcolor{black}{A Comparison of Our Survey With Relevant Surveys
}}
\renewcommand{\arraystretch}{1.5}

%\begin{tabular}{c c l  | c c c c | c c c | c c c c | c c c c | c}
\begin{tabular}{
p{1.2cm}  
p{1cm}  
p{3.2cm}  
| p{0.5cm} p{0.5cm} p{0.5cm} p{0.5cm}  
| p{0.5cm} p{0.5cm} p{0.5cm}           
| p{0.5cm} p{0.5cm} p{0.5cm} p{0.5cm}  
| p{0.6cm} p{0.6cm} p{0.6cm} p{0.6cm}  
| p{1cm}                            
}
\toprule
\multirow{2}{*}{Survey} &
\multirow{2}{*}{Year} &
\multirow{2}{*}{Objective} &
\multicolumn{4}{c|}{Protocol Coverage} &
\multicolumn{3}{c|}{Threat Modeling} &
\multicolumn{4}{c|}{Attack Surfaces} &
\multicolumn{4}{c|}{Qualitative Risk Assessment} &
\multirow{2}{*}{Case Study} \\ \cline{4-18}

&  & &
\tiny MCP & \tiny A2A & \tiny Agora & \tiny ANP &
\tiny Auth. & \tiny SC & \tiny Reli. &
\tiny MCP & \tiny A2A & \tiny Agora & \tiny ANP &
\tiny MCP & \tiny A2A & \tiny Agora & \tiny ANP &
\\
\midrule

\cite{hou2025model} & 2025 & \textcolor{black}{MCP threat taxonomy}  &
\cmark & \xmark & \xmark & \xmark &  \xmark & \xmark & \xmark & \cmark & \xmark & \xmark & \xmark & \xmark & \xmark & \xmark & \xmark & \cmark \\
\midrule

\cite{narajala2025enterprise} & 2025 & \textcolor{black}{MCP security frameworks}  & \cmark & \xmark & \xmark & \xmark & \cmark & \cmark & \cmark & \cmark & \xmark & \xmark & \xmark & \xmark & \xmark & \xmark  &  \xmark & \xmark\\
\midrule

\cite{ehtesham2025survey} & 2025 & \textcolor{black}{Protocol comparison}  & \cmark
 & \cmark & \xmark & \cmark & \cmark & \cmark &  \cmark & \cmark & \cmark & \xmark & \cmark & \xmark & \xmark & \xmark & \xmark & \cmark \\
\midrule

\cite{yang2025iot} & 2025 & \textcolor{black}{MCP-IoT integration}  &
 \cmark & \xmark & \xmark & \xmark &
\xmark & \xmark & \cmark &
\xmark & \xmark & \xmark & \xmark &
\xmark & \xmark & \xmark & \xmark &
\cmark \\
\midrule

\cite{li2025secure} & 2025 & \textcolor{black}{Secure MCP design}  &
 \cmark & \xmark & \xmark & \xmark &
\cmark & \cmark & \xmark &
\cmark & \xmark & \xmark & \xmark &
\xmark & \xmark & \xmark & \xmark &
\cmark \\
\midrule

 \cite{duan2025agent} & 2025 & \textcolor{black}{A2A edge evaluation}  &
\cmark & \cmark & \xmark & \cmark &
\cmark & \xmark & \cmark &
\xmark & \xmark & \xmark & \xmark &
\xmark & \xmark & \xmark & \xmark &
\cmark \\
\midrule

\cite{habler2025building} & 2025 & \textcolor{black}{A2A security analysis}  &
\cmark & \cmark & \xmark & \xmark &
\cmark & \cmark & \cmark &
\cmark & \cmark & \xmark & \xmark &
\xmark & \xmark & \xmark & \xmark &
\cmark \\
\midrule

\cite{louck2025proposal} & 2025 & \textcolor{black}{Secure A2A design }  &
\xmark & \cmark & \xmark & \xmark &
\cmark & \xmark & \cmark &
\xmark & \cmark & \xmark & \xmark &
\xmark & \xmark & \xmark & \xmark &
\cmark \\
\midrule

\cite{he2025comprehensive} & 2025 & \textcolor{black}{Agent threat modeling}  &
\cmark & \cmark & \xmark & \xmark &
\cmark & \cmark & \cmark &
\cmark & \cmark & \xmark & \xmark &
\xmark & \xmark & \xmark & \xmark &
\xmark \\
\midrule

\cite{duan2025ai} & 2025 & \textcolor{black}{Agents in 6G}  &
\cmark & \cmark & \xmark & \xmark &   
\cmark & \xmark & \cmark &           
\xmark & \xmark & \xmark & \xmark &  
\xmark & \xmark & \xmark & \xmark &  
\cmark \\ 
\midrule

\cite{kong2025survey} & 2025 & \textcolor{black}{Protocol security survey} &
\cmark & \cmark & \cmark & \cmark &   
\cmark & \cmark & \cmark &            
\cmark & \cmark & \xmark & \xmark &   
\xmark & \xmark & \xmark & \xmark &   
\cmark \\   
\midrule

\cite{zhang2025survey} & 2025 & \textcolor{black}{IoA collaboration survey}  &
\cmark & \xmark & \cmark & \cmark &   
\xmark & \xmark & \xmark &           
\xmark & \xmark & \xmark & \xmark &  
\xmark & \xmark & \xmark & \xmark &  
\xmark \\ 
\midrule

\cite{wang2025security} & 2025 & \textcolor{black}{IoA security survey}  &
\cmark & \cmark & \cmark & \cmark &
\cmark & \cmark & \cmark &
\cmark & \cmark & \xmark & \xmark &
\xmark & \xmark & \xmark & \xmark &
\xmark \\
\midrule

\cite{wang2025internet} & 2025 & \textcolor{black}{IoA architecture survey}  &
\cmark & \cmark & \cmark & \cmark & 
\cmark & \cmark & \xmark &          
\xmark & \xmark & \xmark & \xmark & 
\xmark & \xmark & \xmark & \xmark & 
\xmark \\ 
\midrule

% \cite{gupta2025ai} & 2025 & ???  &
% \xmark & \xmark & \xmark & \xmark &
% \xmark & \xmark & \xmark &
% \xmark & \xmark & \xmark & \xmark &
% \xmark & \xmark & \xmark & \xmark &
% \cmark \\
% \midrule

\cite{duanagent} & 2025 & \textcolor{black}{Agent communication overview}  &
\cmark & \cmark & \cmark & \cmark &
\cmark & \xmark & \cmark &
\xmark & \xmark & \xmark & \xmark &
\xmark & \xmark & \xmark & \xmark &
\xmark \\
\midrule

\cite{yan2025beyond} & 2025 & \textcolor{black}{Edge IoT Agents}  &
\cmark & \cmark & \xmark & \cmark &
\cmark & \xmark & \cmark &
\xmark & \xmark & \xmark & \xmark &
\xmark & \xmark & \xmark & \xmark &
\xmark \\
\midrule

\cite{sharma2025collaborative} & 2025 & \textcolor{black}{Agent communication risks}  &
\cmark & \cmark & \cmark & \cmark &
\cmark & \xmark & \xmark &
\xmark & \xmark & \xmark & \xmark &
\xmark & \xmark & \xmark & \xmark &
\xmark \\
\midrule

\cite{yang2025survey} & 2025 & \textcolor{black}{Agent Protocol benchmarking}  &
\cmark & \cmark & \cmark & \cmark &   
\cmark & \cmark & \cmark &           
\xmark & \xmark & \xmark & \xmark &  
\xmark & \xmark & \xmark & \xmark &  
\cmark \\
\bottomrule

\textbf{This Paper} & 2026 & Protocol Threat Modeling  &
\cmark & \cmark & \cmark & \cmark &
\cmark & \cmark & \cmark &
\cmark & \cmark & \cmark & \cmark &
\cmark & \cmark & \cmark & \cmark &
\cmark \\
\bottomrule
\end{tabular}
\begin{tablenotes}
          \item \footnotesize \textbf{Note:} MCP: Model Context Protocol; A2A: Agent2Agent protocol; Agora: Agora protocol;
ANP: Agent Network Protocol; Auth: Authentication; SC: Security Concerns; Reli: Reliability; IoA: Internet of Agents.
    \end{tablenotes}
\label{survey_comparison}
\end{table*}

%%%%%%%%%%%%%%%%%%%%%%%
\subsection{Other Protocols and Related Studies}

Existing research on Agora primarily emphasizes communication efficiency and scalability  \cite{marro2024scalable}; the design assumes a cooperative and non-adversarial environment and does not define an explicit threat model. Similarly, ANP is primarily specified through architectural designs and technical specifications that emphasize decentralized identity and interoperability \cite{chang2025agent}. Overall, studies on ANP and Agora currently lack rigorous security assessments, formal threat modeling, and experimental evaluation under adversarial conditions.

Overall, existing studies are fragmented and tend to focus on individual components or isolated interaction patterns, rather than offering a holistic security perspective on the agent communication protocol ecosystem. Kong et al. \cite{kong2025survey} provide a survey for LLM‑driven agent communication and perform experiments using MCP and A2A to illustrate potential vulnerabilities. Zhang et al. \cite{zhang2025survey} review LLM-based multi-agent systems across capabilities, collaboration, architectures, communication, and applications. However, security is treated only at a high level, without formal threat models or protocol-level validation.

In the domain of the Internet of Agents (IoA), Wang et al. \cite{wang2025security} examine identity authentication, cross-agent trust, embodied security, and privacy risks in IoA systems. In a related survey, the same authors \cite{wang2025internet} presented a hierarchical IoA architecture; however, security and privacy are mainly discussed at a conceptual level, without formal threat models or protocol-level enforcement mechanisms. Additional discussions on AI agent communication frameworks in IoT environments can be found in \cite{gupta2025ai} and \cite{duanagent}, though these works do not focus on general-purpose agent communication protocols. Yan et al. \cite{yan2025beyond} explored LLM-based multi-agent systems from a communication-centric perspective and discussed challenges such as efficiency and security; however, security is addressed mainly at a conceptual level, without formal threat models or protocol-level validation. Authors of \cite{sharma2025collaborative} argue that ecosystem fragmentation threatens agentic AI and propose a Web of Agents architecture comprising agent‑to‑agent messaging; however, they overlooked defining threat models and protocol specifications. Yang et al.

Overall, these works highlight the urgency of developing secure, interoperable agent ecosystems but often lack formalized adversary models or empirical validation. Continued research is needed to bridge protocol design with security analysis and practical mitigation strategies.

%%%%%%%%%%%%%%%%%%%%%%%%%%%%%%%%%%%%%%%%%%%%%%%%%%%%%%%%

\section{Foundations of Agent Communication Protocols}
\label{Section_III}

AI assistants have become popular, but are limited by their isolation from data and what they were trained on. Without access to external tools and data sources, they are restricted in providing updated data, acting in the real world, and connecting to external systems. AI agent protocols address this challenge by providing universal materials and interfaces for connecting AI systems with data sources and other AI agents.

\subsection{Model Context Protocol (MCP)}

\begin{figure*}[!t]
    \centering
    \includegraphics[width=\linewidth]{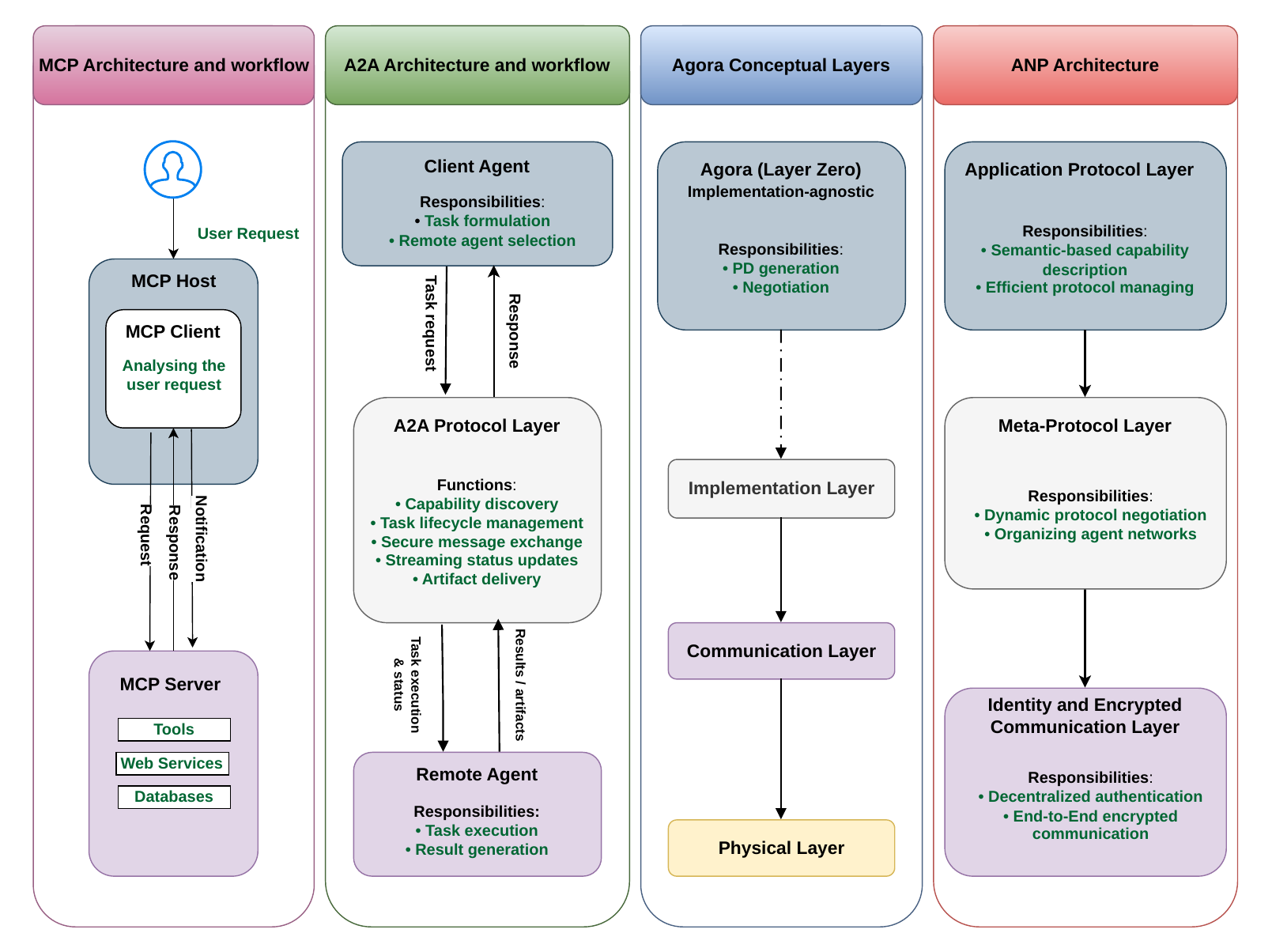}
    \caption{Architecture and workflow of MCP, A2A, Agora and ANP}
    \label{fig:architecture}
\end{figure*}

In 2024, Anthropic introduced MCP as a new open standard protocol for connecting AI agents \cite{anthropic_mcp_2024}. It is open-source and platform-agnostic, enabling agents to have a two-way connection with external tools and facilitating complex workflows. 
As depicted in Figure \ref{fig:architecture}, MCP has three main components, including the MCP host, client, and server. The MCP host is an AI application that hosts the MCP client and offers the environment for executing AI-based operations. The MCP client sends requests to MCP servers and asks about the functionalities that are available, then gets answers about the server's capabilities. 
The MCP server is responsible for managing and providing AI models with resources and has three phases: creation, operation, and update. The creation step is for server registration and configuration. In the operation phase, the MCP server processes requests, invokes tools, handles commands, and enforces a sandbox mechanism. In the last phase, the MCP server maintains security and adaptability by verifying post-update access permissions.
%%%%%%%%%%%%%%%%%NEWWW%%%%%%%%%%%%%%%%%%%%%%%%%%%%%
%%%%%%%%%%%%%%%%%%%%%%%%%%%%%%%%%%%%%%%%%%%%%%

\subsection{Agent2Agent (A2A) Protocol}
In April 2025, Google introduced a new protocol called A2A \cite{rao2025a2a}. The A2A protocol allows AI agents to communicate and exchange information. It is built on existing common standards, including Server-Sent Events (SSE), HTTP(S), and JSON-RPC, and OAuth 2.0 \cite{hardt2012oauth} for mutual agent authentication and access to resources without sharing credentials; it also uses JSON Web Tokens (JWTs) \cite{jones2015json} for compacting and signing tokens.

The A2A protocol provides structured and bidirectional communication between a client agent and a remote agent to distribute task execution. Within this framework, the client agent formulates tasks and transmits them to the remote agent, and then the remote agent works on the tasks to collect information.

Agents advertise their functional capabilities through a standardized agent card in JSON format, and tasks are encapsulated within a well-defined protocol object that has a life cycle; then, the final output is formalized as an artifact. Agents can exchange contextual messages, intermediate responses, artifacts, and user instructions to keep alignment during task execution \cite{a2aproject2025A2A}. By this mechanism, A2A provides a flexible communication platform that supports interoperability and distributed agent collaboration. In Figure \ref{fig:architecture}, the high-level architecture and interaction workflow of A2A is shown.

%%%%%%%%%%%%%%%%%%%%%%%%%%%%%%%%%%%%%%%%%%%%%%%%%%%%%%%%%%%%%%%%%%%%%

\subsection{Agora Protocol}
Agora \cite{marro2024scalable} is an agent communication protocol built to solve the Agent Communication Trilemma in heterogeneous LLM networks. That trilemma captures the fact that versatility (support for diverse message formats and modalities), efficiency (low computational and network costs), and portability (ease of implementation and deployment with minimal human intervention) are inherently incompatible; trying to have a system optimized in all of these dimensions is difficult. Agora uses the capabilities of LLM in natural language comprehension; its main novelty lies in the use of Protocol Documents (PDs) that allow autonomous agents to negotiate, adapt, and modify protocols. Agents share PDs that are a uniquely identified solution with decentralized storage and retrieval and no central authorities, enabling reuse across agents that have never interacted.

Agora has a technology-agnostic design and is a Layer Zero protocol that sits above implementation and communications layers. The meta-protocol layer is responsible for enabling adaptive and negotiable communication through PDs. With this abstraction, Agora separates protocol logic from implementations and makes a technology-agnostic basis for agent communication. Figure \ref{fig:architecture} provides a conceptual layered abstraction of the Agora protocol.

%%%%%%%%%%%%%%%%%%%%%%%%%%%%%%%%%%%%%%%%%%%%%%%%%%%%%%%%%%%%%%%%%%%%

\subsection{Agent Network Protocol (ANP)}
ANP \cite{chang2025agent} \cite{chang2025anp} is an open standard for providing network interoperability between autonomous agents in heterogeneous environments. ANP imagines the IoA as a global, secure, and efficient environment to collaborate with billions of machine entities. The ANP design aims at eliminating data silos, facilitating connectivity between agents, and providing high efficiency in machine-to-machine communication (M2M).

The architecture of the protocol has three layers. The Identity and Encrypted Communication layer integrates the W3C Decentralized Identifiers (DIDs) framework in order to support decentralized authentication, allowing trustless, end-to-end encrypted communication between different agents across different platforms. The Meta-Protocol layer is meant to act as a protocol of protocols where agents may negotiate which communication standard (for example, Agora) to apply to a certain interaction accordingly. Lastly, the Application Protocol Layer defines agent discovery mechanisms, descriptions of capabilities, and execution of tasks in domain-related situations \cite{anp2025github}. 
In Figure \ref{fig:architecture}, the layered architecture of ANP with its identity, meta-protocol, and application layers is demonstrated.

%%%%%%%%%%%%%%%%%%%%%%%%%%%%%%%%%%%%%%%%%%%%%%%
The functional cycle starts with a local agent submitting a standardized search application to a standard discovery process to find a list of available agents. The agent would then access description files about the capabilities of the discovered agents, and what is needed in order to authenticate itself. On this basis, the initiating agent builds and delivers authenticated requests and processes the responses to accomplish collaborative tasks.

%%%%%%%%%%%%%%%%%%%%%%%%%%%%%%%%%%%%%%%%%%%%%%%%%%%%%%%

%%%%%%%%%%%%%%%%%%%%%%%%%%%%%%%%%%%%%%%%%%%%%%%%%%%%%%%%

%%%%%%%%%%%%%%%%%%%%%%%%%%%%%%%%%%%%%%%%%%%%%%%%%%%%%%%%
\section{Threat Model}

\label{Section_IV}

Deploying AI agent communication protocols introduces a wide range of new risks because they allow agents, tools, and external resources to interact in an autonomous and multistep manner. This section presents the reported security threats in AI agent protocols, which are extracted from \cite{yang2025survey, hou2025model, louck2025proposal, biswas2025mcpgovernance,
posta2025mcp_a2a_attacks, wibowo2025toward, bhatt2025etdi, zhao2025mind,
errico2025securing, gaire2025systematization, hasan2025model, he2025automatic,
li2025toward, wang2025mcpguard, shen2026mcp, zhang2025mcp, jamshidi2025securing,
zong2025mcp, song2025beyond, huang2026component, stappen2026agent2agent,
wang2026mcptox, guo2026agent, huang2026model}.

\begin{figure*}[!t]
    \centering
    \includegraphics[width=\linewidth]{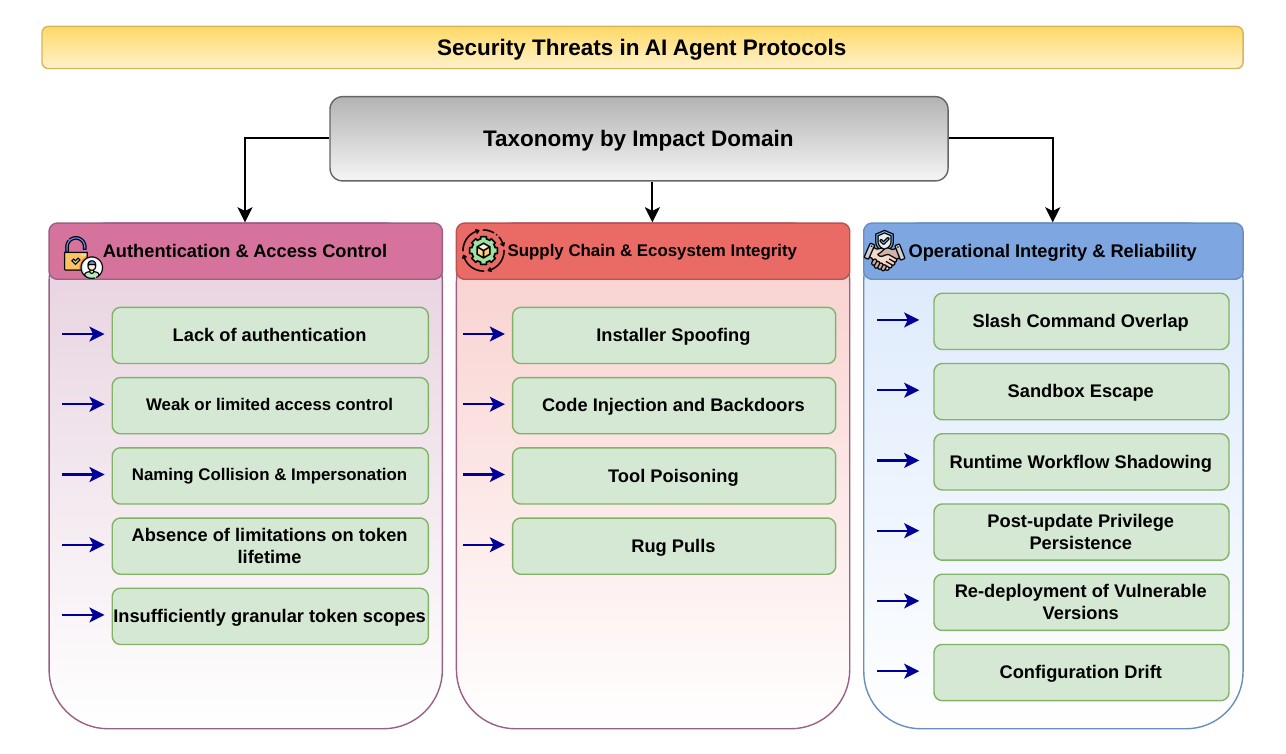}
    \caption{Security threat taxonomy for AI agent communication protocols.}
    \label{fig: Taxonomy}
\end{figure*}

The taxonomy is organized based on impact domains: security threats address authentication \& access control, supply chain \& ecosystem integrity, and operational integrity \& reliability. This structure ensures clarity, avoids redundancy, and aligns with established models such as STRIDE and the CIA triad. The Authentication \& Access Control group of attacks undermines the reliability of agent identity, credential validation, and access enforcement. The Supply Chain \& Ecosystem Integrity group compromises the integrity of AI ecosystem artifacts and update chains. And the Operational Integrity \& Reliability group targets the stable execution and interpretation of tasks across dynamic agent networks, affecting integrity, availability, and coordination reliability. The taxonomy is shown in Figure \ref{fig: Taxonomy}. 
%The summarization of all the attacks is provided in Table \ref{tab:security-threats}.

% %%%%%%%%%%%%%%%%%%%NEWWW%%%%%%%%%%%%%%%%%%%%%%%%%%

%%%%%%%%%%%%%%%%%%%%%%%%%%%%%%%%%%%%%%%%%%%%%

\subsection{Authentication and Access Control}
The authentication and access control group includes attacks that weaken agent identity, credential verification, and access enforcement.

\subsubsection{Lack of authentication}
The early version of MCP did not have authentication mechanisms and was prone to impersonation and spoofing until a later update (MCP v1.2) added token-based authentication that guaranteed identity verification. %\cite{yang2025survey}.

\subsubsection{Weak or limited access control}
Since MCP lacks fine-grained permissions, Access Control List (ACL) granularity is desired on the server side and should be enforced. Coarse permissions fail to add restrictions at the field, endpoint, or task level and expose systems to unpermitted access and privilege escalation. %\cite{yang2025survey}.

\subsubsection{Naming Collision \& Impersonation}
MCP clients discover servers by simply reading their description and name, rather than cryptographic evidence, and there is no central registry to enforce naming rules in decentralized environments. So, in the creation stage of MCP, if a malicious entity registers an MCP server with a name close to a known one, it can impersonate clients. In addition, in open discovery plans of ANP  where capabilities are self-declared, malicious agents can spoof reputable identities or falsify high-value capabilities to attract tasks.

\subsubsection{Absence of limitations on token lifetime}
A2A uses OAuth 2.0 for authentication; however, it does not impose strict expiration durations of tokens for sensitive operations. Consequently, the leaked or intercepted tokens can be stored, allowing the attackers to reuse them for unauthorized access. %\cite{louck2025proposal}.

\subsubsection{Insufficiently Granular Token Scope}
Tokens in the A2A are typically coarse-grained, so they give agents more privileges than they need in their respective work. This coarse scoping leaves a vulnerability to privilege escalation, with all tokens being compromised, allowing attackers to expand their access to areas not originally intended.

\subsection{Supply Chain and Ecosystem Integrity}
This group of attacks focuses on risks related to the integrity of AI ecosystem artifacts and update processes.

\subsubsection{Installer Spoofing}
During the creation phase, attackers are able to publish altered installers or one-click setup programs that install malware or backdoors. Many users have a tendency to use non-official community installers, which do not verify packages or signatures. Attackers can have long-term access, exfiltrate credentials, or reconfigure servers by means of malicious installers. MCP is a much more community-driven system; that means that the attack surface is massive and largely uncontrolled. % \cite{hou2025model}. 

\subsubsection{Code Injection and Backdoors}
MCP servers are often open-source and rely on community-maintained libraries. In addition, malicious or compromised dependencies introduce backdoors into MCP servers. In this regard, without rigorous dependency checks, attackers can slip in code-level exploits, leave behind a backdoor, silently steal data, and increase privileges. Unlike installer spoofing, which is related to the initial stage of the supply chain, this vector concerns the payload/persistence stage after installation. % \cite{hou2025model}. 

\subsubsection{Tool Poisoning}
In the operation phase, tools within the ecosystem can have the same or misleadingly similar names. Besides, MCP enables agents to independently choose the tools according to the names and descriptions. In the case that a malicious tool is given a name or description in order to seem more pertinent, the client may prioritize it. The result is a toolflow hijacking, thereby possibly stealing sensitive data or executing malicious code. % \cite{hou2025model}.

\subsubsection{Rug Pulls}
Rug-pull attacks are a kind of risk in which, at first, adversarial tools or agents act appropriately to gain trust and be integrated into the workflows of critical operations. When the dependency has been established, the malicious party changes their behavior, either removing the desired functionality or adding some bad behavior. MCP and A2A promote dynamic discovery and healthy relationships; therefore, rug pulls are a serious threat to integrity and dependability. Their malicious character lies in the fact that they are activated later, and when they are used, they can bypass the initial checks and use the developed trust relations.% \cite{posta2025mcp_a2a_attacks}.

\subsection{Operational Integrity and Reliability}
This group covers threats that disrupt the stable execution and interpretation of tasks across dynamic agent networks, impacting integrity, availability, and coordination.

\subsubsection{Slash Command Overlap}
The flexible multi-tool environment promoted at MCP provides significant extensibility, yet does not have powerful disambiguation. Several tools define the same or similar slash commands, which threatened the system integrity during the operation phase. In this state, an attacker can insert an incompatible command and cause the MCP client to take unwanted actions (e.g., deleting logs rather than temporary files).

\subsubsection{Sandbox Escape}
MCP is based on local isolation at the operation stage. In case the sandbox implementation contains unpatched vulnerabilities, malicious tools may breach isolation, and then attackers would be able to run arbitrary code on the host, get sensitive system data, or escalate privileges. This is a systemic risk in the enterprise deployment since MCP hosts generally have wide access to tools and resources. % \cite{hou2025model}.

\subsubsection{Runtime Workflow Shadowing}
A less severe form of tool poisoning is the shadowing attack, in which malicious actors pretend to be legitimate tools or agents but shadow these agents during their execution. In this kind of attack, the attacker is already in the path and modifies outputs after a correct tool has been selected. Using this method, the shadowing entity redirects traffic and intercepts workflows, and finally replaces malicious results or modifies outputs. This attack exploits decentralized discovery and can silently corrupt a multi-agent workflow in MCP and A2A protocols. % \cite{posta2025mcp_a2a_attacks}.

\subsubsection{Post-Update Privilege Persistence}
Post-update privilege persistence is a vulnerability that is critical and occurs when outdated privileges or revoked privileges are still valid after an MCP server update. This is a revocation propagation failure related to update time, not a baseline access-control design problem. This may provide adversaries with an opportunity to abuse residual privileges to conduct malicious activities, steal sensitive resources, or cause a system malfunction. % \cite{hou2025model}.

\subsubsection{Re-deployment of Vulnerable Versions}
The probability of redeploying susceptible MCP versions is due to the community-based and decentralized character of the ecosystem. There is no official package management infrastructure or auditing authority to impose the use of secure versions. This leaves a dangerous window between the disclosure of vulnerability and its adoption by a patch system, where attackers can still take advantage of vulnerabilities that are already known. % \cite{hou2025model}. 

\subsubsection{Configuration Drift}
It is defined as the gradual build-up of unwanted configuration drift that goes against a secure configuration. This issue is particularly acute in MCP environments, where end-users localize a server or it is deployed in ecosystems that are community-driven. In a multi-tenant environment, one drift event may accidentally reveal sensitive resources, escalate privileges, or extend access to attackers across tenants. % \cite{hou2025model}. 

This section summarizes security threats reported in the existing literature, which to date focuses primarily on MCP and A2A. Published security analyses for Agora and ANP remain limited, which is expected given that all four protocols are young and still evolving, and that systematic security evaluation has only recently begun to emerge.

%%%%%%%%%%%%%%%%%%%%%%%%%%%%%%%%%%%%%%%%%%%%%%%%%%%%%%%%

% \section{Potential Threats derived from Protocol Design and Architecture}

% \input{Threats Inferred from Protocol Design}

%%%%%%%%%%%%%%%%%%%%%%%%%%%%%%%%%%%%%%%%%%%%%%%%%%%%%%%%

% \section{Proposed Taxonomy}
% \input{texonomy}
%%%%%%%%%%%%%%%%%%%%%%%%%%%%%%%%%%%%%%%%%%%%%%%%%%%%%%%%

\section{Evaluation}
\label{Section_VI}

Risk management in AI agent communication protocols is an important and ongoing process that examines various aspects of security, including identifying, assessing, and mitigating risks. The main goal of risk assessment is to reduce security risks to a manageable level, which is possible by identifying the vulnerabilities that are most likely and impactful. This assessment should examine the entire life cycle of AI agent communication protocols to provide sufficient insight into the risks at each stage of the life cycle.

\subsection{Assessment Methodology}
\begin{figure*}[!t]
    \centering
    \includegraphics[width=\linewidth]{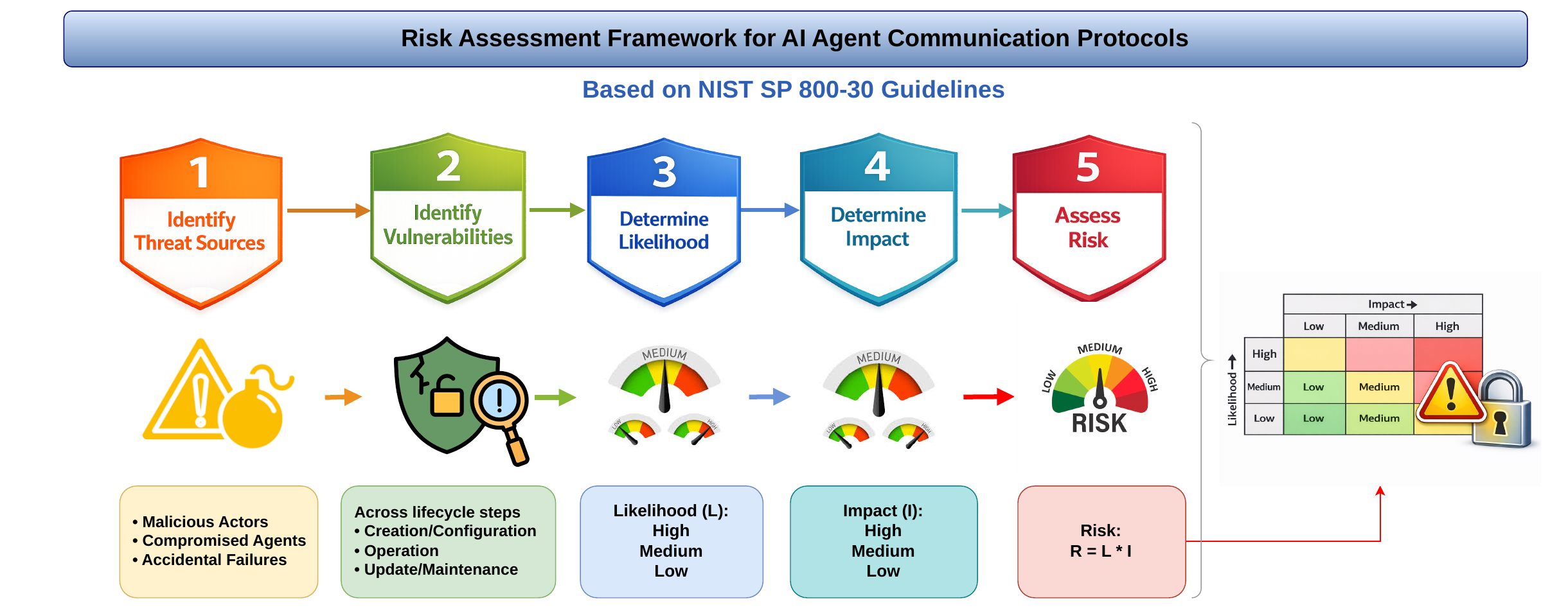}
    \caption{NIST SP 800-30 lifecycle-based risk assessment workflow for AI-agent communication protocols..}
    \label{fig: RiskAssessment}
\end{figure*}

Based on NIST SP 800-30, risk assessment involves five tasks, including (1) identifying threat sources, (2) identifying vulnerabilities that may result from those threat sources and events, (3) determining the likelihood of their occurrence, (4) determining the magnitude of impact for each vulnerability, and finally (5) assessing risk values for the identified threats. Figure \ref{fig: RiskAssessment} shows the risk assessment framework that is adopted in this work. 

% %%%%%%%%%%%%%%%%%%%NEWWW%%%%%%%%%%%%%%%%%%%%%%%%%%

%%%%%%%%%%%%%%%%%%%%%%%%%%%%%%%%%%%%%%%%%%%%%

In this analysis, we employed a qualitative risk assessment approach as the rationale of NIST SP 800-30. All the variables will be rated on a three-level scale, where Low will be rated for rare frequency or minor impact, Medium will be rated for partial and moderate level, and High will be rated for frequent and severe effects. This method is analytical and enables qualitative reasoning across protocols. In order to achieve consistency between various architectures of AI agent communication protocols, qualitative risk levels (high, medium, low) are determined based on protocol-independent reference metrics and not protocol-dependent baselines. 

Since AI agent protocols are still newly proposed with limited public deployments, our evaluation is not based on empirical incident data and reported vulnerabilities. Instead, the ratings of factors are based on a systematic, architecturally based analysis derived from protocol specifications, documented design decisions, and observable security control placement. For each vulnerability, we evaluate the intrinsic exploitability, which is done by assessing the requirement of a security control as mandatory, optional, or non-existent, and the scope and propagation capability allowed by the communication and trust model of the protocol. This method is aligned with design-time risk assessment guidelines in NIST SP 800-30 and ISO/IEC 27005, where risk is estimated in situations of uncertainty based on threat model, expert judgment, and scenario analysis, rather than attack frequency in the past.

\subsubsection{Task 1: Identifying threat sources}
The goal is to identify all potential threat sources and events that could compromise the security of AI agent protocol ecosystems. We have described the sources of threats on an abstract basis, but not on a basis of particular adversary personas. This decision is explained by the fact that we intend to analyze the protocol-level exploitable nature regardless of the sophistication of the attacker.

The origin of threats is classified in three groups that are adopted from NIST SP 800-30, but extended for the  AI agent ecosystem, including (1) malicious human actors who try for unauthorized access or manipulation; (2) compromised agents that were legitimate but have been hacked and used to perform unauthorized actions; and (3) accidental and non-malicious sources such as configuration errors, version skew, dependency drift, and so on.

\subsubsection{Task 2: Identifying vulnerabilities}
The goal of this task is to identify the specific technical vulnerabilities that are created by the threat sources specified in the previous task. In this study, vulnerability is a description of a non-existent, weak, or unenforced security control at the protocol design or deployment level that facilitates a threat event when exploited. The vulnerabilities addressed here were chosen based on them being (i) directly due to protocol specifications or documented design decisions and (ii) controlling risk exposure in at least one stage of the life cycle of the different AI agent communication protocols. The twelve vulnerabilities are categorized based on the stage of the lifecycle they are exposed to:
\begin{itemize}
    \item Creation and Configuration Stage:
    \begin{itemize}
        \item Weak or absent identity verification mechanisms, including shared or static credentials.
        \item Lack of integrity protection for registration artifacts (e.g., unsigned or unverifiable registration files).
        \item Insufficient namespace isolation, enabling impersonation or naming collisions among agents or tools.
        \item Absence of baseline security policy or governance constraints during onboarding.
    \end{itemize}
    \item Operation Stage:
    \begin{itemize}
        \item Lack of mandatory provenance and identity-binding validation, enabling the introduction of unauthorized or untrusted executable components.
        \item Insufficient control over data exchange, leading to context leakage or unauthorized information flow.
        \item Inadequate enforcement of least-privilege principles, allowing privilege escalation via persistent or reused tokens.
        \item Missing rate-limiting, quota enforcement, or backpressure mechanisms, enabling resource exhaustion or denial-of-service conditions.
    \end{itemize}
    \item Update and Maintenance Stage:
    \begin{itemize}
        \item Failure to revoke or reissue credentials after updates, resulting in residual privileges.
        \item Absence of rollback protection or version pinning, enabling downgrade to vulnerable protocol states.
        \item Lack of authentication or integrity verification for maintenance packages or updates.
        \item Uncontrolled transitive dependency evolution, leading to configuration drift or inconsistent security guarantees.
    \end{itemize}
\end{itemize}

\subsubsection{Task 3: Determining the likelihood of occurrence}
This task aims at determining the probability that the vulnerabilities will be used by the threat sources. The intrinsic exploitability and environmental exposure of each protocol in the lifecycle stages are investigated for likelihood estimation. Intrinsic exploitability reflects the design flaws in the protocol, e.g., not expiring tokens and failure to sandbox. On the other hand, environmental exposure shows operational context, e.g., APIs' openness and monitoring procedures. Likelihood ratings represent inherent exploitability at the protocol level within the standard conditions of deployment. This method is compatible with the NIST SP 800-30 qualitative risks framework, where the likelihood is the combination of both design and deployment context.

Table \ref{tab:likelihood} is a qualitative image of the probability of a weakness being utilized during AI agent communication protocols.  When the likelihood is considered to be high, it means that there are conditions that can be readily exploited with few counter-control measures, and when the likelihood is considered to be low, the conditions are mature, and they are well-regulated deployments.

\begin{table*}[t]
%\small
\footnotesize
\centering
\caption{Likelihood Criteria Based on Intrinsic Exploitability and Environmental Exposure}
\label{tab:likelihood}
\renewcommand{\arraystretch}{1.4}
\setlength{\tabcolsep}{4pt}

%\begin{tabular}{|p{2.3cm}|p{7.5cm}|p{7.5cm}|}
% \begin{tabular}{|>{\centering\arraybackslash}m{2cm}|p{7.5cm}|p{7.8cm}|}
\begin{tabular}{|>{\centering\arraybackslash}m{1.9cm}|p{9.4cm}|p{6cm}|}
\hline
\rowcolor{gray!20}
\textbf{Likelihood Level} & \textbf{Intrinsic Exploitability (Design Weakness)} & \textbf{Environmental Exposure (Deployment Context)} \\
\hline

%\rowcolor{red!30}
% \cellcolor{red!30}\textbf{High} &
%\cellcolor{red!30}\parbox[c][\height][c]{\linewidth}{\centering \textbf{High}} &
%\cellcolor{red!30}\parbox[c][1.8cm][c]{\linewidth}{\centering \textbf{High}} &
\cellcolor{red!30}\makebox[\linewidth][c]{\raisebox{-3ex}[0pt][0pt]{\textbf{High}}} &
\textbullet\ Critical protocol design flaws (e.g., no token expiry)\newline
\textbullet\ No sandbox isolation or namespace control\newline
\textbullet\ No provenance or integrity validation
&
\textbullet\ Exploits can be launched remotely via public APIs or open endpoints\newline
\textbullet\ Lack of monitoring or provenance logging\newline
\textbullet\ High exposure to external integrations
\\
\hline

%\rowcolor{yellow!50}
% \cellcolor{yellow!50}\textbf{Medium} &
\cellcolor{yellow!50}\makebox[\linewidth][c]{\raisebox{-3ex}[0pt][0pt]{\textbf{Medium}}} &
\textbullet\ Partial mitigations exist (insufficient to stop attackers)\newline
\textbullet\ Exploitation requires user participation or partial privileges\newline
\textbullet\ Limited enforcement of controls and audits
&
\textbullet\ Logging exists but not enforced\newline
\textbullet\ Semi-trusted or federated environments\newline
\textbullet\ Moderate exposure to external integrations
\\
\hline

%\rowcolor{green!40}
% \cellcolor{green!40}\textbf{Low} &
\cellcolor{green!40}\makebox[\linewidth][c]{\raisebox{-2ex}[0pt][0pt]{\textbf{Low}}} &
\textbullet\ Strong authentication and access control\newline
\textbullet\ Effective sandboxing and isolation\newline
\textbullet\ Enforced least-privilege policies
&
\textbullet\ Minimal or no external API exposure\newline
\textbullet\ Continuous monitoring and logging\newline
\textbullet\ Hardened configuration
\\
\hline

\end{tabular}
\end{table*}

%%%%%%%%%%%%%%%%%%%%%%%%%%%%%%%%%%%%%%%%%%%%%%%%%%%%%%%%%

\subsubsection{Task 4: Determining the magnitude of impact}
Impact level is used to assess the severity of security consequences an attacker can cause by effectively exploiting a vulnerability. We apply a qualitative three-level scale (High, Medium, Low) in line with NIST SP 800-30 \cite{ross2012guide} and \cite{baseri2025evaluation}, modified based on the operational nature of AI agent communication protocols. Each level indicates the potential extent of damage loss to confidentiality, integrity, or availability (CIA) of the agent ecosystem and the intensity of the harm to normal operation. Table \ref{tab:impact-criteria} presents the detailed indicators of using these impact ratings, which brings the qualitative advice of NIST to the security behavior of current AI-agent communication protocols. To improve interpretability, impact is decomposed into (i) CIA degradation and (ii) operational consequences affecting agent ecosystems.

\begin{table*}[t]
%\small
\footnotesize
\centering
\caption{Impact Criteria Based on CIA Degradation and Operational Consequences}
\label{tab:impact-criteria}
\renewcommand{\arraystretch}{1.4}
\setlength{\tabcolsep}{4pt}

%\begin{tabular}{|p{2cm}|p{8.6cm}|p{6.7cm}|}
\begin{tabular}{|>{\centering\arraybackslash}m{1.9cm}|p{9.4cm}|p{6cm}|}
\hline
\rowcolor{gray!20}
\textbf{Impact Level} & \textbf{CIA Impact (Confidentiality, Integrity, Availability)} & \textbf{Operational / System Consequences} \\
\hline

%\rowcolor{red!30}
% \cellcolor{red!30}\textbf{High} &
\cellcolor{red!30}\makebox[\linewidth][c]{\raisebox{-3ex}[0pt][0pt]{\textbf{High}}} &
\textbullet\ Confidentiality breach: large-scale data exfiltration or sensitive context leakage\newline
\textbullet\ Integrity compromise: execution of malicious or unauthorized tasks\newline
\textbullet\ Availability loss: system-wide service disruption or denial-of-service\newline
\textbullet\ Access control failure: unauthorized access due to authentication failure
&
\textbullet\ Unauthorized command execution across agents\newline
\textbullet\ Propagation of poisoned data across ecosystems\newline
\textbullet\ Large-scale financial or operational damage\newline
\textbullet\ Systemic failure affecting multiple entities
\\
\hline

%\rowcolor{yellow!50}
% \cellcolor{yellow!50}\textbf{Medium} &
\cellcolor{yellow!50}\makebox[\linewidth][c]{\raisebox{-4ex}[0pt][0pt]{\textbf{Medium}}} &
\textbullet\ Partial confidentiality exposure: limited data leakage\newline
\textbullet\ Partial integrity degradation: incorrect task execution or routing\newline
\textbullet\ Recoverable integrity violation: reversible state or configuration changes\newline
\textbullet\ Limited availability disruption: recoverable service degradation
&
\textbullet\ Temporary communication or sandbox failures\newline
\textbullet\ Impact limited to a single organization or workflow\newline
\textbullet\ Recoverable system disruption without redesign
\\
\hline

%\rowcolor{green!40}
% \cellcolor{green!40}\textbf{Low} &
\cellcolor{green!40}\makebox[\linewidth][c]{\raisebox{-2ex}[0pt][0pt]{\textbf{Low}}} &
\textbullet\ No significant confidentiality loss or data compromise\newline
\textbullet\ Minimal integrity impact\newline
\textbullet\ Negligible or short-term availability impact
&
\textbullet\ Minor configuration errors or transient failures\newline
\textbullet\ Localized incident with negligible business impact\newline
\textbullet\ Immediate recovery via rollback or patching
\\
\hline

\end{tabular}
\end{table*}

%%%%%%%%%%%%%%%%%%%%%%%%%%%%%%%%%%%%%%%%%%%%%%%%%%%%%%%%%%%%%%%%%%%

\subsubsection{Task 5: Assessing Risk}

The last stage of the risk assessment process is to calculate the security risk of each vulnerability that was identified during the previous tasks. This work is in accordance with  NIST SP 800-30 \cite{ross2012guide} and conceptually aligns with ISO/IEC 27005:2022 \cite{iso27005_2022}. The total risk value is calculated by:
\begin{equation}
\label{eq:risk-formula}
R = L \times I
\end{equation}
where R is the overall risk, L is the probability of a threatening event, and I is defined as the magnitude of impact. The formula is popular in cybersecurity evaluation systems (e.g., \cite{baseri2025evaluation}), which has been applied here to evaluate risks in the communication protocols of AI agents. Table \ref{tab:risk-matrix} is the resulting matrix that gives a comprehensive method for comparing security risk among protocols and lifecycle phases and shows how overall risk severity is determined by the interaction between likelihood and impact.

For each vulnerability, we assign $L$ and $I$ on a three-level ordinal scale (Low = 1, Medium = 2, High = 3) using the criteria in Tables~\ref{tab:likelihood}, \ref{tab:impact-criteria}. Because the ordinal product $R = L \times I$ is used only to support consistent lookup in the qualitative matrix (Table~\ref{tab:risk-matrix}) and not as a standalone numeric "risk level", we map scores to final risk categories exactly as the matrix specifies: Low Risk for $R \in \{1,2\}$, Medium Risk for $R \in \{3,4\}$, and High Risk for $R \in \{6,9\}$.

\begin{table*}[t]
\small
\centering
\caption{Qualitative Risk Matrix for AI Agent Protocols (Based on $R = L \times I$).}
\label{tab:risk-matrix}
\renewcommand{\arraystretch}{1.5}
\begin{tabular}{|p{2cm}|>{\columncolor{yellow!50}}p{4.8cm}|>{\columncolor{red!30}}p{4.8cm}|>{\columncolor{red!30}}p{4.8cm}|}
\hline
\rowcolor{gray!20}
\footnotesize
 \textbf{Likelihood →/ Impact ↓ } & \footnotesize
 \textbf{Low} & \footnotesize
 \textbf{Medium} & \footnotesize
 \textbf{High} \\
\hline

\footnotesize
\cellcolor{gray!20}\textbf{High} &
\scriptsize Limited exposure but significant impact.& \scriptsize 
Medium exposure and serious damage. & \scriptsize 
Catastrophic system failure. \\
\hline

\footnotesize
 \cellcolor{gray!20} \textbf{Medium} &
\scriptsize  \cellcolor{green!40} Unlikely events and moderately impactful. & \scriptsize 
\cellcolor{yellow!50} Moderate likelihood and impact. & \scriptsize 
\cellcolor{red!30}Frequent vulnerability with tangible impact. \\
\hline

\footnotesize
 \cellcolor{gray!20} \textbf{Low} &
\scriptsize  \cellcolor{green!40} Minimal exposure with limited consequence. & \scriptsize 
\cellcolor{green!40} Moderate exposure with minor effect. & \scriptsize 
\cellcolor{yellow!50}High exposure but low consequence. \\
\hline
\end{tabular}

\newcommand{\riskbox}[2]{\colorbox{#1}{\strut\ #2\ }}
\begin{tablenotes}
\item \footnotesize \textbf{Note:} 
\riskbox{green!40}{Low Risk}; 
\riskbox{yellow!50}{Medium Risk}; 
\riskbox{red!30}{High Risk}
\end{tablenotes}

\end{table*}

\subsection{Lifecycle-Based Evaluation Framework}
Despite the fact that individual operational phases, including creation, operation, and update, are only formally specified in MCP, all agent communication protocols tend to act through similar security-relevant states. This study uses a generalized three-phase life cycle model to be able to do inter-protocol comparisons with other models. The justification is consistent with the NIST SP 800-30 risk assessment process, which identifies threats and vulnerabilities during the system development and maintenance process.

By this framework, the activities of any protocol are mapped to similar functions in the life cycle: Creation/configuration involves identity registration, capability discovery, and initiation. Operations contain runtime data exchange and the invocation of a tool. Update/maintenance includes patching, protocol negotiation, and version control.

\subsubsection{\textbf{Stage 1: Creation/configuration}}

The creation and configuration phase is the initial point when an AI agent or tool enters a multi-agent ecosystem. Identity establishment, registration integrity, and initial trust anchors are specified here, and most security vulnerabilities start then, since these safeguards cannot come into effect before this point. During this stage, we analytically compare MCP, A2A, Agora, and ANP about how they manage identity validation, component registration, integrity verification, and namespace governance, and represent them in Table \ref{tab:stage1-final}.

\begin{table*}[htbp]
%\begin{table*}[t]
\centering
\small
\renewcommand{\arraystretch}{1.3}
\caption{Risk Assessment for the Creation/Configuration Stage}
\label{tab:stage1-final}

\begin{tabular}{|>{\centering\arraybackslash}m{3cm}|p{0.8cm}|p{4.7cm}|p{4.75cm}|p{0.05cm}|p{0.05cm}|p{0.05cm}|p{2cm}|}

\hline
\rowcolor{gray!20}
\footnotesize \textbf{Vulnerability} & \footnotesize \textbf{Protocol} & \footnotesize \textbf{Likelihood Basis} & \footnotesize \textbf{Impact Basis} & \footnotesize \textbf{L} & \footnotesize \textbf{I} & \footnotesize \textbf{R} & \footnotesize \textbf{Key Reason} \\

\hline

%-------------------------
% \footnotesize \multirow{4}{3.2cm}{\textbf{Weak or absent identity verification mechanism}}
\footnotesize \multirow{4}{3.2cm}{\parbox[c]{3.2cm}{\centering \vspace{0.3em}\textbf{Weak or absent identity verification mechanism}}}
& \footnotesize MCP & \footnotesize Agents and tools can register without strict identity validation, making impersonation highly probable. & \footnotesize An impersonated client or server directly compromises the trust chain, so it allows complete impersonation of trusted tools or agents, compromising the system’s trust chain. & \cellcolor{red!30} & \cellcolor{red!30} & \cellcolor{red!30} & \footnotesize No strict identity validation \\

& \footnotesize A2A & \footnotesize Authentication is based on OAuth2+JWT. Risk arises only if the scopes are too broad or tokens are long-lived. & \footnotesize Bad scopes or leaked tokens let an attacker submit unauthorized requests, but damage is usually local thanks to scoped tokens. & \cellcolor{yellow!50} & \cellcolor{yellow!50} & \cellcolor{yellow!50} & \footnotesize OAuth2/JWT limits misuse \\

& \footnotesize Agora & \footnotesize Agents declare their identities themselves, and identity proof is optional; there is little control against imitation. & \footnotesize A forged identity can mislead multiple agents and redirect data flows. & \cellcolor{red!30} & \cellcolor{red!30} & \cellcolor{red!30} & \footnotesize Self-declared identities \\

& \footnotesize ANP & \footnotesize ANP integrates W3C DID and E2E encryption, providing strong identity assurance. & \footnotesize Strong W3C DID and E2E encryption prevent impersonation; compromise would be local only. & \cellcolor{green!40} & \cellcolor{yellow!50} & \cellcolor{green!40} &  \footnotesize DID + E2E identity binding \\
\hline

%-------------------------
% \footnotesize \multirow{4}{3.2cm}{\textbf{Lack of integrity protection for registration artifacts}}
\footnotesize \multirow{4}{3.2cm}{\parbox[c]{3.2cm}{\centering \vspace{0.3em}\textbf{Lack of integrity protection for registration artifacts}}}

& \footnotesize MCP   & \footnotesize  New components can be added without obligatory verification, providing an open window for malicious code or fake configurations. & \footnotesize Malicious code executes in a trusted runtime; total integrity failure. & \cellcolor{red!30} & \cellcolor{red!30} & \cellcolor{red!30} & \footnotesize Unsigned artifacts allow injection \\

& \footnotesize A2A   & \footnotesize The A2A agent card and negotiation parameters are not cryptographically signed. Integrity depends on the deployment environment, and in weak configurations, tampering becomes more likely. & \footnotesize Integrity failures affect only the specific agent instance relying on an altered agent card. Verification is deployment-dependent, so it usually does not compromise the entire network. & \cellcolor{yellow!50} & \cellcolor{yellow!50} & \cellcolor{yellow!50} & \footnotesize Integrity depends on deployment \\

& \footnotesize Agora & \footnotesize Negotiation details between agents can be changed without detection. & \footnotesize Altered files can change how agents communicate or execute. & \cellcolor{red!30} & \cellcolor{red!30} & \cellcolor{red!30} & \footnotesize Negotiation data can be altered \\

& \footnotesize ANP & \footnotesize Most implementations support verification, but the process is manual and incomplete. & \footnotesize The meta-protocol layer validates negotiations and testing. Incomplete checks allow hidden tampering across organizations. & \cellcolor{yellow!50} & \cellcolor{yellow!50} & \cellcolor{yellow!50} & \footnotesize Partial verification support \\
\hline

%-------------------------
% \footnotesize \multirow{4}{3.2cm}{\textbf{Insufficient namespace isolation}}
\footnotesize \multirow{4}{3.2cm}{\parbox[c]{3.2cm}{\centering \vspace{0.3em}\textbf{Insufficient namespace isolation}}}

& \footnotesize MCP & \footnotesize Public registry is possible without uniqueness enforcement, which enables spoofing. & \footnotesize Leads to full spoofing and potential credential theft. Enables attackers to intercept or impersonate legitimate agents. & \cellcolor{red!30} & \cellcolor{red!30} & \cellcolor{red!30} & \footnotesize No uniqueness enforcement \\

& \footnotesize A2A & \footnotesize Agent identity is self-declared using the Agent Card with no global uniqueness enforcement. Collisions across deployments are possible. & \footnotesize A wrong agent card could misroute tasks, but mutual OAuth prevents total network takeover and collision remains local to each environment. & \cellcolor{yellow!50} & \cellcolor{yellow!50} & \cellcolor{yellow!50} & \footnotesize No global identity uniqueness \\

& \footnotesize Agora & \footnotesize Agents are self-named; duplication is not prevented. & Trust relationships were fully compromised at the negotiation. & \footnotesize \cellcolor{red!30} & \cellcolor{red!30} & \cellcolor{red!30} & \footnotesize Self-naming causes collisions \\

& \footnotesize ANP & \footnotesize DIDs inherently ensure uniqueness. & \footnotesize Since DIDs ensure uniqueness, there is almost no naming-based failure. & \cellcolor{green!40} & \cellcolor{green!40} & \cellcolor{green!40} & \footnotesize DID ensures uniqueness \\
\hline

%-------------------------
% \footnotesize \multirow{4}{3.2cm}{\textbf{Absence of baseline security policy or governance constraints}}
\footnotesize \multirow{4}{3.2cm}{\parbox[c]{3.2cm}{\centering \vspace{0.3em}\textbf{Absence of baseline security policy or governance constraints}}}

& \footnotesize MCP & \footnotesize Tool Developers have the ability to publish new components without an approval workflow and no community review, but numerous servers are managed locally, minimizing global exposure. & \footnotesize  Weak governance leads to slower detection of malicious components. & \cellcolor{yellow!50} & \cellcolor{yellow!50} & \cellcolor{yellow!50} & \footnotesize No approval workflow \\

& \footnotesize A2A & \footnotesize The review is based on the deploying organization. Inconsistent pre-deployment review makes the problem moderately likely. & \footnotesize Misconfigurations degrade reliability but rarely cause full compromise. & \cellcolor{yellow!50} & \cellcolor{yellow!50} & \cellcolor{yellow!50} & \footnotesize Deployment-dependent review \\

& \footnotesize Agora & \footnotesize Community-based moderation exists, but it is inconsistent. & \footnotesize Policy ambiguity mostly causes interoperability errors, not total failure. & \cellcolor{yellow!50} & \cellcolor{yellow!50} & \cellcolor{yellow!50} & \footnotesize Inconsistent moderation \\

& \footnotesize ANP & \footnotesize Multi-organization contributes with limited oversight. & \footnotesize Cross-org miscoordination could slow incident response but not cause a full breach. & \cellcolor{yellow!50} & \cellcolor{yellow!50} & \cellcolor{yellow!50} & \footnotesize Cross-org governance gaps \\
\hline

\end{tabular}

\vspace{-0.5em}
\newcommand{\riskbox}[2]{\colorbox{#1}{\strut\ #2\ }}

\begin{flushleft}
\footnotesize\textit{\textbf{Note:} L: Likelihood level, I: Impact level, R: Risk level.}

%\vspace{0.2em}
\footnotesize
\riskbox{green!40}{Low};
\riskbox{yellow!50}{Medium};
\riskbox{red!30}{High}
\end{flushleft}

\end{table*}

In the four representative protocols, the most security-important phase of the protocol lifecycle is the creation and configuration phase, since it determines under what underlying trust relations, what identity relationships, and what integrity guarantees that future interactions will rely on. The likelihood, impact, and risk assessments (Table \ref{tab:stage1-final}) are consistent in indicating that the vulnerabilities in this stage are highly exploitable with predominantly high consequences, with only some exceptions in ANP.

The results can be explained by three general patterns: (1) If the enrollment paths are open or weakly authenticated, the risk of information disclosure increases, as seen in the MCP and Agora protocols. An attacker can impersonate trusted components or inject malicious logic before any safeguards are activated. (2) A2A uses OAuth2/JWT authentication and mutual token validation, which reduces the risk of abuse compared to MCP and Agora, yet it does not have an established pre-deployment integrity global registry to enforce uniqueness, placing A2A in the middle of the risk spectrum. (3) ANP is the only protocol that cryptographically guarantees identity. Overall, ANP is characterized by low-median risk, as opposed to a consistent benefit across all vulnerabilities. In general, the risk analysis shows that there is no protocol with a low risk level for all the vulnerabilities in this stage. These results affirm that the creation/configuration phase is the most impactful period of the lifecycle, as early weaknesses carry on to become systemic and hard to fix in the future.

%%%%%%%%%%%%%%%%%%%%%%%%%%%%%%%%%%%%%%%%%%%%%%%%%%%%%%%

\subsubsection{\textbf{Stage 2: Operation}}
The operation phase is where agents are communicating with each other, calling routines, and performing actions. In this stage, the vulnerabilities are the dynamic interaction threats, i.e., unsafe data exchanges, logic manipulation, and so on. As this stage includes continuous communication and real-time decision-making, the vulnerabilities in runtime isolation, encryption, or authorization directly impact agent ecosystems. This discussion demonstrates that the architectural decisions influence runtime security exposure in agent networks.

\begin{table*}[htbp]
%\begin{table*}[t]
\centering
\small
\renewcommand{\arraystretch}{1.3}
\caption{Risk Assessment for the Operation Stage}
\label{tab:stage2-final}

\begin{tabular}{|>{\centering\arraybackslash}m{3cm}|p{0.8cm}|p{4.7cm}|p{4.75cm}|p{0.05cm}|p{0.05cm}|p{0.05cm}|p{2cm}|}
\hline
\rowcolor{gray!20}
\footnotesize \textbf{Vulnerability} & \footnotesize \textbf{Protocol} & \footnotesize \textbf{Likelihood Basis} & \footnotesize \textbf{Impact Basis} & \footnotesize \textbf{L} & \footnotesize \textbf{I} & \footnotesize \textbf{R} & \footnotesize \textbf{Key Reason} \\
\hline

% =============================
% \footnotesize \multirow{4}{3cm}{\textbf{Lack of provenance and identity-binding validation}}
\footnotesize \multirow{4}{3.2cm}{\parbox[c]{3.2cm}{\centering \vspace{0.3em}\textbf{Lack of provenance and identity-binding validation}}}

& \footnotesize MCP
& \footnotesize No mandatory signed tool manifest or provider-bound tool ID; colliding tools can be accepted.
& \footnotesize Wrong-provider execution or poisoned tools can lead to systemic integrity loss.
& \cellcolor{red!30} & \cellcolor{red!30} & \cellcolor{red!30}
& \footnotesize No binding between tool and provider \\ 

& \footnotesize A2A
& \footnotesize OAuth2/JWT authenticates transport, but agent-card/task claims lack mandatory issuer-bound provenance.
& \footnotesize Integrity impact is localized to affected agents, but unverified claims can misroute tasks.
& \cellcolor{yellow!50} & \cellcolor{yellow!50} & \cellcolor{yellow!50}
& \footnotesize lack of mandatory issuer-bound provenance \\

& \footnotesize Agora
& \footnotesize PD/routine negotiation lacks mandatory integrity enforcement; manipulated PDs can be introduced in negotiation.
& \footnotesize Manipulated PDs propagate via negotiation, causing system-wide workflow corruption.
& \cellcolor{red!30} & \cellcolor{red!30} & \cellcolor{red!30}
& \footnotesize PD integrity not enforced \\

& \footnotesize ANP
& \footnotesize DID identity is anchored, but provenance of application-layer dependencies may be bypassed in implementation.
& \footnotesize Compromised logic or dependencies can propagate across layers despite identity anchoring.
& \cellcolor{yellow!50} & \cellcolor{red!30} & \cellcolor{red!30}
& \footnotesize Dependency provenance gap \\
\hline

% =============================
% \footnotesize \multirow{4}{3cm}{\textbf{Insufficient control over data exchange}}
\footnotesize \multirow{4}{3.2cm}{\parbox[c]{3.2cm}{\centering \vspace{0.3em}\textbf{Insufficient control over data exchange}}}

& \footnotesize MCP
& \footnotesize Agents share the whole context with tools.
& \footnotesize Complete user context exposure leads to system-wide confidentiality breach.
& \cellcolor{red!30} & \cellcolor{red!30} & \cellcolor{red!30}
& \footnotesize Full context exposure \\

& \footnotesize A2A
& \footnotesize Message structure is mediated, but semantic minimization is not enforced. Leakage could occur in case an agent overshares.
& \footnotesize Only certain parts of messages that an agent sends are leaked; A2A does not automatically carry along context.
& \cellcolor{yellow!50} & \cellcolor{yellow!50} & \cellcolor{yellow!50}
& \footnotesize Limited structure control \\

& \footnotesize Agora
& \footnotesize No strict separation of private/public channels; NL negotiation leaks metadata.
& \footnotesize NL-based negotiation and PD sharing expose reasoning and metadata, affecting integrity and confidentiality.
& \cellcolor{red!30} & \cellcolor{red!30} & \cellcolor{red!30}
& \footnotesize Metadata leakage via negotiation \\

& \footnotesize ANP
& \footnotesize Context encryption is not mandatory; metadata may be exposed by policy.
& \footnotesize DID-based communication restricts leakage to the layer with weak policy.
& \cellcolor{yellow!50} & \cellcolor{yellow!50} & \cellcolor{yellow!50}
& \footnotesize Layer-bounded exposure \\
\hline

% =============================
% \footnotesize \multirow{4}{3cm}{\textbf{Inadequate enforcement of least-privilege principles}}
\footnotesize \multirow{4}{3.2cm}{\parbox[c]{3.2cm}{\centering \vspace{0.3em}\textbf{Inadequate enforcement of least-privilege principles}}}

& \footnotesize MCP
& \footnotesize Tokens rarely expire and can be reused by compromised components.
& \footnotesize Persistent tokens enable the attackers to give tool commands
as the legitimate agent and give complete administrative control.
& \cellcolor{red!30} & \cellcolor{red!30} & \cellcolor{red!30}
& \footnotesize Long-lived tokens enable takeover \\

& \footnotesize A2A
& \footnotesize Long-lived tokens and lack of rotation increase privilege reuse.
& \footnotesize Unauthorized access persists, but scoped tokens limit escalation.
& \cellcolor{yellow!50} & \cellcolor{yellow!50} & \cellcolor{yellow!50}
& \footnotesize Scoped tokens reduce elevation impact \\

& \footnotesize Agora
& \footnotesize Temporary privileges in negotiation may persist longer than intended.
& \footnotesize Temporary privilege leakage enables lateral access but not full control.
& \cellcolor{yellow!50} & \cellcolor{yellow!50} & \cellcolor{yellow!50}
& \footnotesize Temporary privilege persistence \\

& \footnotesize ANP
& \footnotesize Layered roles limit escalation, but cross-layer token reuse is possible.
& \footnotesize When one of the layers is compromised, cross-layer identity may be used to compromise core routing and service orchestration
& \cellcolor{yellow!50} & \cellcolor{red!30} & \cellcolor{red!30}
& \footnotesize Cross-layer privilege reuse \\
\hline

% =============================
% \footnotesize \multirow{4}{3cm}{\textbf{Missing rate-limiting, quota enforcement, or backpressure mechanisms}}
\footnotesize \multirow{4}{3.2cm}{\parbox[c]{3.2cm}{\centering \vspace{0.3em}\textbf{Missing rate-limiting, quota enforcement, or backpressure mechanisms}}}

& \footnotesize MCP
& \footnotesize Sandbox limits full DoS, but loops in agent requests remain possible.
& \footnotesize Availability degradation remains local to the affected server due to the sandbox.
& \cellcolor{yellow!50} & \cellcolor{yellow!50} & \cellcolor{yellow!50}
& \footnotesize Sandbox limits damage\\

& \footnotesize A2A
& \footnotesize unlimited message queues and real-time streaming with
SSE allow flooding.
& \footnotesize DoS flooding can disable coordination between agents.
& \cellcolor{red!30} & \cellcolor{red!30} & \cellcolor{red!30}
& \footnotesize JSON-RPC/SSE flooding \\

& \footnotesize Agora
& \footnotesize Dynamic composition can cause unexpected cost/runtime overload.
& \footnotesize Dynamic routine generation increases latency but does not crash
the ecosystem.
& \cellcolor{red!30} & \cellcolor{yellow!50} & \cellcolor{red!30}
& \footnotesize Runtime overload risk \\

& \footnotesize ANP
& \footnotesize Cross-layer messaging adds overhead but avoids collapse.
& \footnotesize DoS affects the delay in message transmissions between layers; however, routing is not shut down.
& \cellcolor{yellow!50} & \cellcolor{yellow!50} & \cellcolor{yellow!50}
& \footnotesize Cross-layer messaging prevents outages \\
\hline

\end{tabular}

\vspace{-0.5em}
\newcommand{\riskbox}[2]{\colorbox{#1}{\strut\ #2\ }}

\begin{flushleft}
\footnotesize\textit{\textbf{Note:} L: Likelihood level, I: Impact level, R: Risk level.}

%\vspace{0.2em}
\footnotesize
\riskbox{green!40}{Low};
\riskbox{yellow!50}{Medium};
\riskbox{red!30}{High}
\end{flushleft}

\end{table*}

In the operation stage, agents actively run tasks, share context, negotiate workflow, and, on top of all that, rely on long-lived credentials. Because of the nature of this stage, the four protocols have a high level of security exposure. In this stage, operational risk is introduced by runtime behaviors, such as dynamic message processing, privilege propagation, or coordination between different agents. The assessment results for this phase, shown in Table \ref{tab:stage2-final}, indicate that among the vulnerabilities assessed, MCP and Agora have a high overall risk, due to the lack of runtime code-integrity enforcement in MCP and to dynamic negotiation based on PD in Agora. A2A presents a moderate risk across all vectors because OAuth2/JWT reduces exploitability and message-level mediation limits impact, but the downside of this protocol at this stage is that it does not provide any guarantees for semantic validation or strict token lifetime management. ANP exhibits mixed behavior; it has strong DID-based authentication, but the multi-layered dependencies cause the impact to increase once an attack happens. Overall, the results obtained for the operation phase show that although runtime flexibility is essential for agent collaboration, it is the main security risk factor in all protocols.

%%%%%%%%%%%%%%%%%%%%%%%%%%%%%%%%%%%%%%%%%%%%%%%%%%%%%%%
%%%%%%%%%%%%%%%%%%%%%%%%%%%%%%%%%%%%%%%%%%%%%%%%%%%%%%%
%%%%%%%%%%%%%%%%%%%%%%%%%%%%%%%%%%%%%%%%%%%%%%%%%%%%%%%

\subsubsection{\textbf{Stage 3: Update \& Maintenance}}

The update and maintenance phase is the most crucial stage in the life cycle of AI agent communication protocols since it determines how agents evolve, acquire new capabilities, remove outdated components, and sustain long-term trust. Any weaknesses in version verification, dependency integrity, or rollback mechanisms can introduce persistent and systemic vulnerabilities that spread throughout entire multi-agent ecosystems. Therefore, this section aims to evaluate how these protocols handle updates and dependency changes. The findings are shown in Table \ref{tab:stage3-final}.

\begin{table*}[htbp]
%\begin{table*}[t]
\centering
\small
\renewcommand{\arraystretch}{1.3}
\caption{Consolidated Risk Assessment for the Update/Maintenance Stage}
\label{tab:stage3-final}

\begin{tabular}{|>{\centering\arraybackslash}m{3cm}|p{0.8cm}|p{4.7cm}|p{4.75cm}|p{0.05cm}|p{0.05cm}|p{0.05cm}|p{2cm}|}
\hline
\rowcolor{gray!20}
\footnotesize \textbf{Vulnerability} & \footnotesize \textbf{Protocol} & \footnotesize \textbf{Likelihood Basis} & \footnotesize \textbf{Impact Basis} & \footnotesize \textbf{L} & \footnotesize \textbf{I} & \footnotesize \textbf{R} & \footnotesize \textbf{Key Risk Note} \\
\hline

%-------------------------
% \footnotesize \multirow{4}{3cm}{\textbf{Failure to revoke or reissue credentials after updates}}
\footnotesize \multirow{4}{3.2cm}{\parbox[c]{3.2cm}{\centering \vspace{0.3em}\textbf{Failure to revoke or reissue credentials after update}}}

& \footnotesize MCP
& \footnotesize No token revocation enforcement or re-authentication cycles after updates.
& \footnotesize Obsolete tokens give attackers full access and complete compromise of tools and data.
& \cellcolor{red!30} & \cellcolor{red!30} & \cellcolor{red!30}
& \footnotesize No forced revocation after updates \\

& \footnotesize A2A
& \footnotesize Although OAuth2/JWT supports rotation, there is no mandatory re-auth after capability changes.
& \footnotesize OAuth2/JWT limits token misuse to the compromised agent.
& \cellcolor{yellow!50} & \cellcolor{yellow!50} & \cellcolor{yellow!50}
& \footnotesize OAuth2 supports rotation but not enforced post-update \\

& \footnotesize Agora
& \footnotesize PD-based workflows can create roles that persist unless the agent
explicitly revokes them. There is no ”session reset” after PD updates.
& \footnotesize Malicious actors use remaining delegated roles to impersonate real peers and disrupt the network.
& \cellcolor{yellow!50} & \cellcolor{red!30} & \cellcolor{red!30}
& \footnotesize Stale delegated roles persist \\

& \footnotesize ANP
& \footnotesize Although DID/E2E secures identity, application-layer credentials can remain valid after an update.
& \footnotesize Remaining credentials allow temporary privilege retention, but DID/E2E limits access and threat scope.
& \cellcolor{yellow!50} & \cellcolor{yellow!50} & \cellcolor{yellow!50}
& \footnotesize DID/E2E limits stale-credential scope \\
\hline

%-------------------------
% \footnotesize \multirow{4}{3cm}{\textbf{Absence of rollback protection or version pinning}}
\footnotesize \multirow{4}{3.2cm}{\parbox[c]{3.2cm}{\centering \vspace{0.3em}\textbf{Absence of rollback protection or version pinning}}}

& \footnotesize MCP
& \footnotesize Lack of version pinning or mandatory signing makes it easy to redeploy older server versions.
& \footnotesize Downgrade restarts patched vulnerabilities and allows data corruption in production.
& \cellcolor{red!30} & \cellcolor{red!30} & \cellcolor{red!30}
& \footnotesize No version pinning or mandatory signing \\

& \footnotesize A2A
& \footnotesize A2A is stateless and backward-compatible; rollback must be done explicitly.
& \footnotesize Downgrade reintroduces old agent-side logic errors, but OAuth2 authentication prevents network takeover.
& \cellcolor{yellow!50} & \cellcolor{yellow!50} & \cellcolor{yellow!50}
& \footnotesize Stateless design \\

& \footnotesize Agora
& \footnotesize PD negotiation allows agents to accept prior PD hashes; no global governance.
& \footnotesize Outdated PDs bring agents to insecure communication, compromising network trust.
& \cellcolor{red!30} & \cellcolor{red!30} & \cellcolor{red!30}
& \footnotesize Outdated PDs weaken network semantics \\

& \footnotesize ANP
& \footnotesize ANP provides version hints, but enforcement is optional.
& \footnotesize Biased degradation within organizations does not destroy negotiation but undermines it.
& \cellcolor{yellow!50} & \cellcolor{yellow!50} & \cellcolor{yellow!50}
& \footnotesize Optional version enforcement \\
\hline

%-------------------------
% \footnotesize \multirow{4}{3cm}{\textbf{Lack of authentication or verification for maintenance packages or updates}}
\footnotesize \multirow{4}{3.2cm}{\parbox[c]{3.2cm}{\centering \vspace{0.3em}\textbf{Lack of authentication or verification for maintenance packages or updates}}}

& \footnotesize MCP
& \footnotesize Lack of obligatory signing for post-deployment tool updates makes poisoned updates plausible.
& \footnotesize Unsigned updates allow attackers to place arbitrary code with full server privileges.
& \cellcolor{red!30} & \cellcolor{red!30} & \cellcolor{red!30}
& \footnotesize Unsigned updates enable arbitrary code injection \\

& \footnotesize A2A
& \footnotesize A2A does not force verification; mitigation is partial and implementation-dependent.
& \footnotesize No specification on the signing of agent-side code or scripts. Unsigned updates corrupt the task-handling logic of one agent, but OAuth2 ensures the attack does not propagate further.
& \cellcolor{yellow!50} & \cellcolor{yellow!50} & \cellcolor{yellow!50}
& \footnotesize Verification is optional and deployment-dependent \\

& \footnotesize Agora
& \footnotesize Since decentralized references distribute PDs and there is no end-to-end signature validation, spoofed PD updates are likely.
& \footnotesize Spoofed PDs propagate false logic through the execution of many agents globally.
& \cellcolor{red!30} & \cellcolor{red!30} & \cellcolor{red!30}
& \footnotesize Spoofed PD updates can corrupt workflows globally \\

& \footnotesize ANP
& \footnotesize Although DIDs sign core artifacts, transitive dependencies at the application layer may bypass checks.
& \footnotesize Unsigned packages affect one service layer; DID-based signing limits spread and prevent full network corruption.
& \cellcolor{yellow!50} & \cellcolor{yellow!50} & \cellcolor{yellow!50}
& \footnotesize Transitive dependencies remain a gap \\
\hline

%-------------------------
% \footnotesize \multirow{4}{3cm}{\textbf{Uncontrolled transitive dependency evolution}}
\footnotesize \multirow{4}{3.2cm}{\parbox[c]{3.2cm}{\centering \vspace{0.3em}\textbf{Uncontrolled transitive dependency evolution}}}

& \footnotesize MCP
& \footnotesize Resources evolve without synchronization; heterogeneous server versions desynchronize capabilities.
& \footnotesize Version drift causes incompatibility but does not result in total service loss.
& \cellcolor{red!30} & \cellcolor{yellow!50} & \cellcolor{red!30}
& \footnotesize Asynchronous evolution breaks compatibility \\

& \footnotesize A2A
& \footnotesize Federated agents upgrade independently; message compatibility limits breakage, but drift is common.
& \footnotesize Agent upgrades are asynchronous and lead to temporary message errors; OAuth2 limits this to local workflow failures.
& \cellcolor{yellow!50} & \cellcolor{yellow!50} & \cellcolor{yellow!50}
& \footnotesize Independent upgrades cause recoverable inconsistencies \\

& \footnotesize Agora
& \footnotesize Decentralized PD evolution probably leads to divergence until consensus forms.
& \footnotesize Divergent PD evolution violates negotiation semantics and may end multi-agent collaboration.
& \cellcolor{red!30} & \cellcolor{red!30} & \cellcolor{red!30}
& \footnotesize PD divergence can halt collaboration \\

& \footnotesize ANP
& \footnotesize Layered design decreases divergence, but cross-layer mismatches remain in multi-organization ecosystems.
& \footnotesize Cross-layer cadence incompatibility slows down communication; the self-healing meta-protocol handles it across cycles.
& \cellcolor{yellow!50} & \cellcolor{yellow!50} & \cellcolor{yellow!50}
& \footnotesize Cross-layer drift mitigated by layered architecture \\
\hline

\end{tabular}

\vspace{-0.5em}
\newcommand{\riskbox}[2]{\colorbox{#1}{\strut\ #2\ }}

\begin{flushleft}
\footnotesize\textit{\textbf{Note:} L: Likelihood level, I: Impact level, R: Risk level.}

%\vspace{0.2em}
\footnotesize
\riskbox{green!40}{Low};
\riskbox{yellow!50}{Medium};
\riskbox{red!30}{High}
\end{flushleft}

\end{table*}

The assessment of the update \& maintenance phase reveals that all four protocols are at medium to high risk, and the reason for this trend is the reintroduction of trust dependencies after deployment. MCP presents a high risk level; the lack of forced revocation, version pinning, and post-deployment signing are the main reasons for this high risk level, which makes rollback attacks, poisoned updates, and transitive drift possible. Agora also has a high risk due to decentralized PD propagation, as old or fake PDs can spread throughout the network and disrupt the negotiation. In the case of A2A, OAuth2/JWT restricts privilege propagation and limits the risk radius, but its stateless backward compatibility still remains a problem and permits local compromise. ANP is relatively less risky due to its DID/E2E identity anchor and layered signing; however, inter-organizational asynchrony and mismatched transitive dependencies still cause moderate operational disruption. Overall, the update \& maintenance phase shows a critical systemic pattern: when agents enter long-term operations, the lack of coordinated revocation and version management increases risk across protocols.

%%%%%%%%%%%%%%%%%%%%%%%%%%%%%%%%%%%%%%
Taken together, our assessment of the three phases of the protocol lifecycle shows that all of the protocols examined still present security and integrity risks that cannot be ignored. Each protocol exhibits strengths in some areas, but none of them offers complete protection across the entire lifecycle. If these protocols are to be deployed in environments with sensitive data or requiring cross-domain coordination, they must address their fundamental vulnerabilities.

% \FloatBarrier

%%%%%%%%%%%%%%%%%%%%%%%%%%%%%%%%%%%%%%%%%%%%%%%%%%%%%%%%

%%%%%%%%%%%%%%%%%%%%%%%%%%%%%%%%%%%%%%%%%%
\FloatBarrier
\section{Experimental Case Study}

\label{case_study}

In this section, we conduct an empirical security evaluation focused on the MCP protocol to convert one of the previously hypothesized risks into a falsifiable claim. This focus is intentional: MCP uniquely standardizes agent-to-tool/server invocation, whereas A2A, Agora, and ANP primarily define inter-agent communication. As a result, the “wrong-provider tool execution” scenario examined here is not directly applicable to those protocols without redefining the system under test (SUT), attack surface, and success criteria.

To rigorously evaluate this MCP-specific risk, we implement a minimal yet realistic SUT and perform controlled, repeatable experiments with auditable evidence to demonstrate whether the identified weakness can be exploited in practice. The objective is not a speculative critique, but a measurement-driven validation that a concrete security invariant can be violated under realistic deployment assumptions.

The evaluation follows a measurement-driven methodology based on repeated trials, explicit success criteria, and rate-based reporting. This approach aligns with prior rigorous studies in AI agent security \cite{fu2024imprompter}, \cite{zou2025poisonedrag}, \cite{deng2024pandora}, \cite{chen2025struq}, which emphasize quantifying attack success rather than relying solely on conceptual analysis.

MCP does not standardize a protocol-level mechanism that uniquely and cryptographically binds a tool’s identity to its provider when multiple MCP servers are concurrently available. Therefore, tool identity is effectively resolved using non-unique identifiers (tool name and description) and client-side heuristics, which introduces an operation-stage tool identity ambiguity titled "Lack of mandatory provenance and identity binding validation". When two servers advertise the same tool name, the client must resolve the collision via a selection policy. Under plausible resolver behaviors, this ambiguity can lead to tool/provider misbinding, in which an invocation intended for a legitimate provider is executed on an unintended or malicious provider. 

A scenario is assumed to show this high-impact weakness in MCP. Consider a financial institution that deploys an MCP-integrated AI assistant to support merchant payments. The assistant invokes MCP tools to retrieve merchant details and initiate payment authorization requests via approved payment service providers. The orchestrator gets a user request such as “Authorize a merchant payment via the official bank-approved verified PSP tool.” It then selects a tool advertised by one or more MCP servers and calls it.

Two MCP servers are configured in the client environment; one of them is a legitimate payment service provider exposing the tool \texttt{payments.authorize\_transaction} for the bank-approved PSP, and the other one is a malicious MCP server exposing a tool with an identical name that returns valid-looking “approved” responses and emits a distinct evidence marker upon execution. For any invocation targeting a tool from the intended provider, the system must never execute that tool from the attacker provider, even when a colliding tool is present. A single observed violation of this invariant under the stated threat model constitutes evidence that the hypothesized weakness is real in the studied setting.

MCP client resolves tools based on name and description without a cryptographic binding to the provider identity. When the AI agent initiates a payment authorization, the orchestrator misbinds the tool invocation to the malicious server instead of the legitimate one. As a result, the authorization request is sent to an attacker-controlled endpoint, leading to integrity and confidentiality violations in a financial workflow.

A strict threat model is adopted to avoid overclaiming; the attacker has the capability to deploy an MCP server and make it visible through the same discovery surface as legitimate servers, or it can be a previously legitimate one that has added new tools. The attacker does not have the capability to compromise hosts, break cryptography, or perform a network Man-in-the-Middle (MITM) attack. The attacker's goal is to cause an invocation intended for a legitimate tool to execute on the attacker's server.

\begin{figure}[!t]
    \centering
    \includegraphics[width=3.2 in]{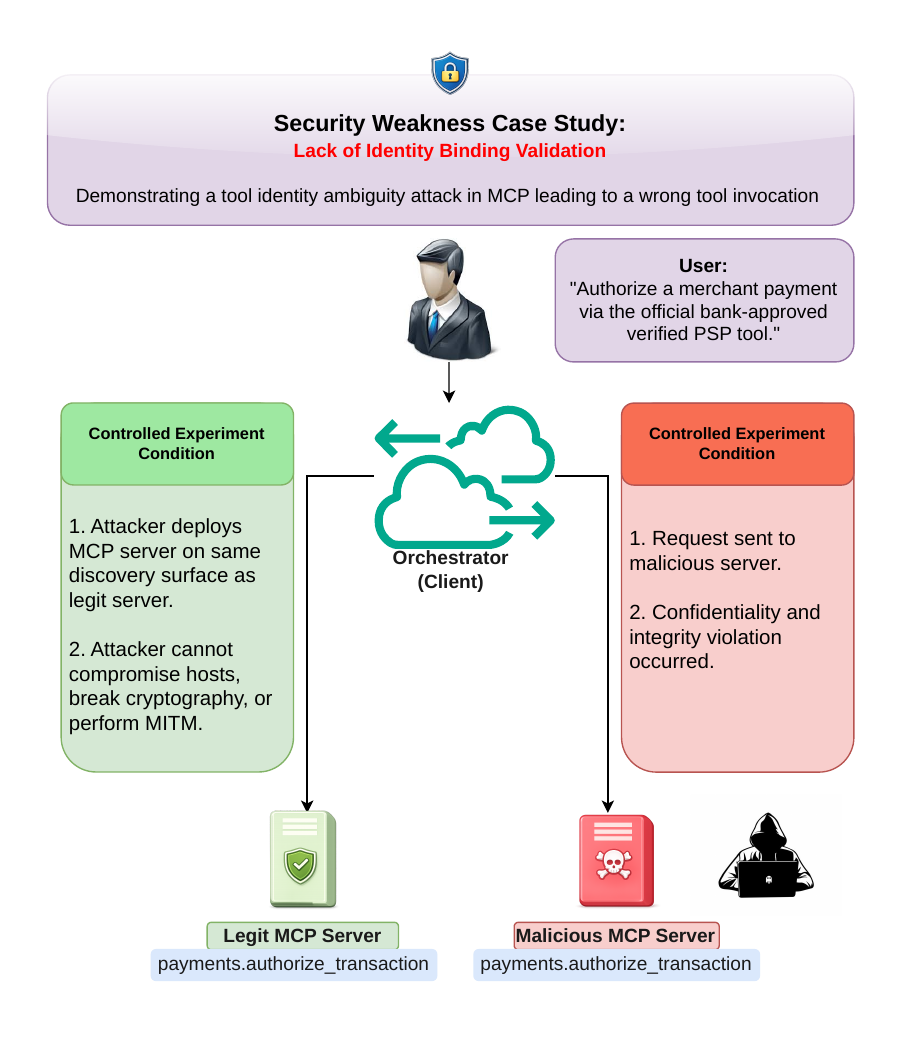}
    \caption{Experimental Architecture for MCP Tool Identity Ambiguity Evaluation.}
    \label{fig: testbed}
\end{figure}

SUT consists of an orchestrator (MCP client runtime) that performs tool discovery, selection, and invocation, and two MCP servers, including a legitimate one and an attacker's, with at least one colliding tool. The experimental testbed architecture is presented in Figure \ref{fig: testbed}. The payment task is kept constant across trials, and we vary only the minimal factors needed for causality, including the presence or absence of the attacker server, selection policy (ordering-first, best-match scoring, or random tie-break on equal scores), and metadata equivalence.

MCP discovery is typically out-of-band, such as static configuration, local provisioning, or an external registry. This experiment models the practice explicitly via a registry-like directory that contains JSON server specifications (server ID, command, and arguments). The runner loads these specifications, and the orchestrator connects to each server over MCP (JSON-RPC over stdio), issues an initialization, and requests tool advertisements. The resulting tool inventory is cached in a candidate index used during the trial loop. It is essential to note that this experiment is not focused on dynamic discovery bugs; it is about what happens after multiple servers are simultaneously available, which is exactly where binding matters.

%%%%%%%%%%%%%%%%%%%%%%%%%%%%%%%%%%%%%%%%%%%%%%%%%
\begin{table*}[t]
\centering
% \small
\footnotesize
\caption{Experiments Demonstrating Unauthorized Tool Execution Due to Missing Identity Binding Validation in MCP}
%\resizebox{\textwidth}{!}{%
\begin{tabular}{|p{0.4cm}|p{1.8cm}|p{3.5cm}|p{4.8cm}|p{1.7cm}|p{0.5cm}|p{1.4cm}|p{0.7cm}|}
\hline
\rowcolor[HTML]{C0C0C0} \textbf{Exp} & \textbf{Discovery surface} & \textbf{Resolver policy} & \textbf{Attacker lever / manipulation} & \textbf{Tie rule} & \textbf{N} & \textbf{\#Violation} & \textbf{VR}\\
\hline
\multirow{2}{*}{A} & static multi-server configuration & First-match (pick first provider advertising tool) & Legit listed before attacker & N/A & 100 & 0 & 0.000\\
 &  &  & Attacker listed before legit & N/A & 100 & 100 & 1.000\\
\hline
\rowcolor[HTML]{F0F0F0} \multirow{2}{*}{B} & registry directory & First-match & Registry filename order favors legit & N/A & 100 & 0 & 0.000\\
\rowcolor[HTML]{F0F0F0} &  &  & Registry filename order favors attacker & N/A & 100 & 100 & 1.000\\
\hline
\multirow{3}{*}{C} & registry directory & Best-match (task-to-metadata scoring) & Legit has metadata/``trust cues'' & Deterministic (no tie) & 100 & 0 & 0.000\\
&  &  & Attacker has metadata/``trust cues'' & Deterministic (no tie) & 100 & 100 & 1.000\\
&  &  & Metadata cloned $\Rightarrow$ tie regime & Random tie-break & 100 & 52 & 0.520\\
\hline
\end{tabular}%
\begin{tablenotes}
          \item \footnotesize \textbf{Note:} Exp: Experiment, N: Number of trials, VR: Violation Rate.
    \end{tablenotes}
\label{tab:seven-experiments}
\end{table*}

%%%%%%%%%%%%%%%%%%%%%%%%%%%%%%%%%%%%%%%%%%%%%%%%%%%%%%%

MCP v1.25.0 (the latest version at the time of writing) is used in this experiment to ensure that the measured behavior is not a legacy artifact. We also set deterministic seeds for randomized tie-breaking where needed and record (i) the discovery artifacts (tool lists), (ii) selection evidence (scores/ties), and (iii) the invoked server per trial, enabling independent replay and verification of results.

Inspired by the attack success rate (ASR), which is used in recent AI security evaluations such as \cite{zou2025poisonedrag} and \cite{chen2025struq}, we define a new, clear success metric where success is counted over repeated trials as Violation Rate (VR), measured using a binary event. A trial is considered a violation if the client invokes \texttt{authorize\_payment} on the attacker provider while the legitimate server is available.

Violation Rate (VR) is reported as

\begin{equation}
VR = \frac{\#\text{violations}}{N}
\label{eq:violation-rate}
\end{equation}

Where N is the number of trials. In this setting, “attack success” is precisely the misbinding event (wrong server executed), making VR an ASR-style measure specialized to server-tool binding failures.

In the experiments pipeline, the orchestrator (MCP client) establishes a session with each server and performs initialization and tool list to build a unified tool candidate index across servers, and then executes the constant payment task repeatedly. At each trial, the orchestrator performs several tasks, including selection, invocation, and evidence logging. The selection is to choose among candidate tools with the same name using an agent-like best-match score derived from the user task and tool metadata; if candidates are indistinguishable, it applies a random tie-break.

We evaluate the hypothesis across multiple experiments. This evaluation does not depend on a misconfigured endpoint or an accidental registry defect. The most notable finding across all experiments is that when several MCP servers are available, the client must perform cross-server tool resolution. Since the tool name is not globally unique and there is no mechanism for binding tools to servers in the face of multiple available tools with the same name, the system enters an ambiguity regime where the orchestrator is able to pick an unwanted provider under realistic resolution policies such as ordering, heuristic best-match, or randomized tie-break.

We demonstrate this ambiguity under three increasingly stronger conditions: (A) explicit multi-server static discovery, (B) registry-style discovery, in which ordering is caused by realistic listing and filenames, and (C) a best-match selection that is agent-like, in which identically similar metadata creates stochastic misbinding. Table \ref{tab:seven-experiments} summarizes the conditions of different experiments. Collectively, these findings indicate that the bug is not an isolated scoring bug, nor is it a collusion discovery-surface bug; it is a system properties issue of ambiguous tool identity under collisions.

In all experiments, when identical tools are present, but the identity of the provider is not tied to the tool identity, a non-zero VR can be observed. Deterministic first-match resolution generates VR=1.0 when an attacker is visible first, and metadata-based best-match selection generates VR=1.0 with an attacker injecting trust cues into the tool description. Importantly, with the attacker cloning metadata to cause indistinguishable candidates, when we eliminate ordering bias by randomizing tie-breaking, misbinding occurs with VR=0.52 in 100 trials. This shows that the failure is not due to a specific scoring function but rather to a lack of resolution in ambiguity between tools and providers. A proposed approach to address this issue is to make the tool identity provider-dependent and validated by supported cryptographic certificates/signatures, which we will address in future work.

%%%%%%%%%%%%%%%%%%%%%%%%%%%%%%%%%%%%%%%%%%
\FloatBarrier
\section{Conclusion}
\label{Conclusion}

This paper presents the first systematic and focused review of the security of emerging AI agent communication protocols at a time when the pace of industry adoption is much faster than the security maturity of the ecosystem. By bringing together scattered initial results and developing a standard taxonomy of security threats, we show that these protocols share common structural weaknesses in authentication, supply chain integrity, operational reliability, and so on. Specific vulnerabilities for MCP and A2A have already been documented in the literature; however, for Agora and ANP, no security analysis has been published to date. However, through architectural reasoning, we identify a diverse set of previously unreported attack vectors that are based on the nature of their decentralized trust models and natural language-based interoperability.

In order to provide a more practical foundation for secure deployment, we proposed a risk assessment model of a lifecycle based on NIST and assessed threats in the different stages of protocols. Our findings indicate that these protocols are not only subjected to individual attacks but also have some systemic weaknesses due to undeveloped governance, inconsistent identity assumptions, and cross-protocol collaboration. Importantly, we complement the qualitative analysis with a measurement-driven case study on MCP that instantiates an operation stage weakness: when tool identity is not cryptographically bound to provider identity, cross-server tool collisions can yield wrong-provider tool execution under realistic resolver policies. By quantifying this misbinding as a violation rate across controlled trials, we demonstrate how a design-level ambiguity can translate into a concrete, reproducible security failure. These risks are becoming more consequential as AI agents expand to enterprise workflows, multi-vendor environments, and even financial operations. Our analysis tries to inform future researchers by predicting both established and emerging threats in order to develop strong, security-sensitive agentic ecologies.

%%%%%%%%%%%%%%%%%%%%%%%%%%%%%%%%%%%%%%%%%%
\FloatBarrier
\section{Future Research Directions}
\label{Future_work}

Even though this research offers one of the first in-depth analyses of security in the emerging protocols of AI agents, a number of key research gaps are still available. The first gap is related to providing a special security layer for MCP. MCP is the only protocol that has been structured for AI agent-tool communication, but it suffers from the lack of a comprehensive security layer to protect it from many attacks. These are especially challenging for financial or safety-critical processes, where the invocation of the tool may cause a sensitive or irreversible activity. For our future work, we would like to implement a formally defined security extension to MCP that includes cryptographic identity anchoring, ephemeral access credentials, and verifiable permission scoping, making MCP suitable enough to be used in enterprise-grade and regulated settings. These additions introduce overhead, so feasibility must be validated through measurements of latency/throughput impact and failure behavior under partial adoption.

Till now, all the research discussed individual protocol security threats, but in this work, we have demonstrated that MCP, A2A, ANP, and Agora use completely different trust assumptions, authentication, and validation strategies, and this creates a chance of confusion, downgrade, and relay-abuse attacks when they are combined. So, cross-protocol security standards and interoperability hardening are urgently required. Any interoperability layer must define a minimal canonical mapping (identity + capability + provenance) and include explicit binding to protocol context to mitigate relay and downgrade paths.

%%%%%%%%%%%%%%%%%%%%%%%%%%%%%%%%%%%%%%%%%%
\FloatBarrier
\section*{Acknowledgements}
\textcolor{black}{The authors express their gratitude to the anonymous reviewers for their valuable feedback. Additionally, the authors sincerely appreciate the support received from the Canadian Institute for Cybersecurity (CIC).}

%\section*{Declaration of generative AI}

%\textcolor{red}{The authors used ChatGPT (OpenAI) solely to improve the language and clarity of the manuscript. The authors reviewed and edited the resulting text as needed and take full responsibility for the content of the publication.}

% \begin{thebibliography}{1}
%\section{References}

% \bibliographystyle{IEEEtran}
% \bibliography{References}
\makeatletter
\def\bibfont{\footnotesize}
\makeatother
\setlength{\bibsep}{0pt}

\FloatBarrier
\bibliographystyle{elsarticle-num}
\bibliography{References1}

\end{document}